\documentclass[11pt,a4paper]{article} 
\usepackage[utf8]{inputenc}
\usepackage{bm}
\usepackage{float}
\usepackage{amsmath}
\usepackage{amsfonts}
\usepackage{amssymb}
\usepackage{cite}
\usepackage{hyperref}
\usepackage[capitalise,noabbrev]{cleveref}
\usepackage[titletoc]{appendix}
\usepackage{graphicx}
\usepackage{multirow}
\usepackage{physics}
\everymath={\displaystyle}
\crefrangelabelformat{equation}{(#3#1#4)--(#5#2#6)}
\crefname{equation}{}{}
\crefname{figure}{Figure}{Figures}
\crefname{tabular}{Table}{Tables}
\crefname{section}{Section}{Sections}
\newcommand\numberthis{\refstepcounter{equation}\tag{\theequation}}
\newcommand{\Ast}{\mathop{\scalebox{1.5}{\raisebox{-0.2ex}{$\ast$}}}}%
\newcommand{\pw}{\mathbb{P}}
\newcommand{\cw}{\mathbb{W}}
\newcommand{\cd}{\mathbb{D}}
\newcommand{\n}{\bm{n}}
\newcommand{\dpp}{d^{+}}
\newcommand{\dn}{d^{-}}
\newcommand{\dpn}{d^{\pm}}
\newcommand{\Lp}{L^{+}}
\newcommand{\Ln}{L^{-}}
\newcommand{\unitx}{\bm{\hat{x}}}

\newcommand{\unitz}{\bm{\hat{z}}}
\newcommand{\betaL}{\beta^{-}}
\newcommand{\betaR}{\beta^{+}}
\newcommand{\betaLR}{\beta^{\pm}}
\newcommand{\nuGN}{\bm{\nu}_{\rm GN}}
\newcommand{\nuNS}{\bm{\nu}_{\rm NS}}
\newcommand{\delt}{\delta_\theta}
\newcommand{\delh}{\delta_h}
\newcommand{\deldp}{\delta_{d^{+}}}
\newcommand{\deldn}{\delta_{d^{-}}}
\newcommand{\tx}{\theta_x}
\newcommand{\tz}{\theta_z}
\newcommand{\Dgn}{\Delta_{\rm GN}}
\newcommand{\Dns}{\Delta_{\rm NS}}
\newcommand{\Sn}{S_{\rm N}}
\newcommand{\Si}{S_{\rm I}}
\newcommand{\Earea}{\mathcal{C}_{\rm area}}
\newcommand{\Ebulk}{E_{\rm bulk}}
\newcommand{\Egn}{E_{\rm GN}}
\newcommand{\Egs}{E_{\rm GS}}
\newcommand{\Ens}{E_{\rm NS}}
\newcommand{\wgn}{\omega_{\rm GN}}
\newcommand{\wgs}{\omega_{\rm GS}}
\newcommand{\wns}{\omega_{\rm NS}}
\newcommand{\psig}{\psi_{\rm g}}
\newcommand{\Sgn}{\Gamma_{\rm GN}}
\newcommand{\Sgs}{\Gamma_{\rm GS}}
\newcommand{\Sns}{\Gamma_{\rm NS}}
\newcommand{\cgn}{C_{\rm GN}}

\newcommand{\cns}{C_{\rm NS}}
\newcommand{\gin}{\gamma_{\rm IN}}

\newcommand{\ggi}{\gamma_{\rm GI}}
\newcommand{\gis}{\gamma_{\rm IS}}
\newcommand{\ggn}{\gamma_{\rm GN}}
\newcommand{\ggs}{\gamma_{\rm GS}}
\newcommand{\gns}{\gamma_{\rm NS}}

\newcommand{\etal}{{\it et al.\ }}
\newcommand{\ie}{i.e.\ }
\newcommand{\eg}{e.g.\ }
\begin{document}

\begin{center}
\textbf{\huge Young and Young--Laplace equations for a static ridge of nematic liquid crystal, and transitions between equilibrium states} 
\end{center}
\noindent
{\Large J. R. L. Cousins$^{1,2}$\footnote{joseph.cousins@strath.ac.uk / joseph.cousins@glasgow.ac.uk}, B. R. Duffy$^{1}$\footnote{b.r.duffy@strath.ac.uk}, S. K. Wilson$^{1}$\footnote{s.k.wilson@strath.ac.uk}, N. J. Mottram$^{2}$\footnote{nigel.mottram@glasgow.ac.uk}} \\
\noindent
{\small ${}^{1}$Department of Mathematics and Statistics, University of Strathclyde, Livingstone Tower, 26 Richmond Street, Glasgow G1 1XH, United Kingdom} \\
{\small ${}^{2}$School of Mathematics and Statistics, University of Glasgow, University Place, Glasgow G12 8QQ, United Kingdom} \\
{\small(Dated: 15th November 2021)} 

\begin{abstract}
Motivated by the need for greater understanding of systems that involve interfaces between a nematic liquid crystal, a solid substrate, and a passive gas that includes nematic--substrate--gas three-phase contact lines, we analyse a two-dimensional static ridge of nematic resting on a solid substrate in an atmosphere of passive gas.
Specifically, we obtain the first complete theoretical description for this system, including nematic Young and Young--Laplace equations,
and then, under the assumption that anchoring breaking occurs in regions adjacent to the contact lines, we use the nematic Young equations to determine the continuous and discontinuous transitions that occur between the equilibrium states of complete wetting, partial wetting, and complete dewetting.
In particular, in addition to continuous transitions analogous to those that occur in the classical case of an isotropic liquid, we find a variety of discontinuous transitions, as well as contact-angle hysteresis, and regions of parameter space in which there exist multiple partial wetting states that do not occur in the classical case.
\end{abstract}

\section{Introduction}
\label{sec:Intro}

For the past 50 years or so, technological interest in liquid 
crystals has largely been focused on the visual display 
market, where Liquid Crystal Displays (LCDs) are still the dominant technology \cite{HANDBOOK}. In recent years, however, the push to exploit the optical, dielectric, and viscoelastic anisotropies of liquid crystals has led to the development of devices used in medicine, flow processing, microelectronic production, and adaptive-lens technologies \cite{Gentili2012,Sengupta2013,Algorri2019,Kim2017,Zou2018}.
These devices often involve liquid crystal droplets or films, which are complicated multiphase systems that involve interfaces between the liquid crystal, a solid substrate, and a passive gas, and often include liquid crystal--substrate--gas three-phase contact lines.
Theoretical studies of liquid crystal droplets or films often use theories of wetting and dewetting for isotropic droplets and films which do not account for the full anisotropic nature of liquid crystals \cite{SONINBOOK,LamCummings2020,LamCummings2018,LinCummings2013,LinCummings2013B,Braun2000,Vandenbrouck1999,Manyuhina2014}.

\subsection{Wetting and dewetting phenomena}
\label{sec:intro:wetting}

Simply stated, \emph{wetting} and \emph{dewetting} are the phenomena in which a liquid advances and retreats, respectively, over a substrate \cite{wetBOOK}.
When a finite volume of liquid advances or retreats over a flat horizontal substrate, it will eventually reach an equilibrium state.
This equilibrium state is known as: the \emph{complete wetting} state (sometimes also called the perfectly wetting state), which we denote by $\cw$, when the liquid completely coats the substrate; the \emph{complete dewetting} state, which we denote by $\cd$, when the substrate completely repels the liquid; and the \emph{partial wetting} state, which we denote by $\pw$, when the liquid partially coats the substrate.
Transitions between these equilibrium states can occur as a result of changes in the liquid or substrate material properties (due to, for example, changes in temperature) that cause the liquid to advance or retreat over the substrate and/or change its contact angle.
The classification of the equilibrium states and the transitions between them is well known for an isotropic liquid \cite{wetBOOK}.

Wetting and dewetting phenomena have been of scientific interest for centuries, and are now of increasing technological importance \cite{Extrand2016}. 
For systems in which creating a uniform liquid film (\ie complete wetting) is required, wetting is essential and dewetting is undesirable \cite{wetBOOK}. 
However, in other situations, dewetting can be desirable, and can be initiated in a variety of ways, such as amplification of thermal fluctuations on the liquid free surface, nucleation at impurities, chemical treatment of the substrate, and non-uniform evaporation \cite{Demirel1999}. In recent years there has been considerable research in the area of tailored dewetting of liquid films to produce patterned films \cite{Gentili2012,Bramble2010,Zou2018}. The thermal, mechanical, and chemical stability of liquid films is therefore an area of considerable research effort, and understanding and controlling the onset of dewetting is crucial for creating and maintaining both uniform and patterned films. 

\subsection{Wetting and dewetting phenomena for liquid crystals}

For liquid crystals, which are anisotropic liquids that typically consist of either rod-like or disc-like molecules that tend to align locally to minimise molecular interaction energies, wetting and dewetting phenomena can be more complicated than they are for isotropic liquids. 
The local orientational order of liquid crystal molecules allows for a mathematical description of the average molecular orientation of the liquid crystal in terms of a unit vector called the director $\n$ \cite{ISBOOK}.
As well as an orientational order, many liquid crystal phases also possess positional order; for example, smectic liquid crystals (smectics) self-organise into two-dimensional layers, and this positional ordering may affect the wetting and dewetting behaviour \cite{Ostrovski2001}. 
However, in the present work we consider only {\it nematic liquid crystals} (nematics), which possess orientational but not positional ordering.

A variety of effects, including spinodal dewetting and nucleation at impurities \cite{LamCummings2018,Braun2000,vanEffenterre2003}, can cause the dewetting of nematic films.
In particular, such dewetting can involve competition between many effects, including internal elastic forces, alignment forces on the interfaces, gravity, van der Waals forces, and in cases in which an external electromagnetic field is applied, electromagnetic forces \cite{Vix2000}.
Many experimental studies have considered delicate balances between a number of these effects in different situations, for instance close to the isotropic--nematic phase transition \cite{Herminghaus1998,Vandenbrouck1999,Ravi2015}, near to a contact line \cite{Poulard2005,Delabre2009,ReyHerreraValencia2014}, or in the presence of an external electromagnetic field \cite{YokoyamaKobayashiKamei1985,Schafferetal2000,Oswald2010a}. 
Since in the present work we consider length-scales greater than a nanometre-scale, it is appropriate to neglect van der Waals forces \cite{wetBOOK} and we consider only the competition between elastic forces, alignment forces on the interfaces, and gravity.

\subsection{Liquid crystal anchoring}

As mentioned above, the alignment forces on the interfaces between the gas and the nematic (the gas--nematic interface) and the nematic and the substrate (the nematic--substrate interface) can play an important role in wetting and dewetting behaviour \cite{vanderWielen2000}. 
The dependence of these interactions on the orientational anisotropy typically results in energetic preferences for the director to align either normally or tangentially to the interfaces, which leads to interfacial energies that are anisotropic; this is known as \emph{weak anchoring}.
An energetic preference for the director to align normally to an interface is known as \emph{weak homeotropic anchoring}, and an energetic preference for the director to align tangentially to an interface is known as \emph{weak planar anchoring}.
The strength of the energetic preference for a homeotropic or planar alignment of the director on an interface is measured by a parameter called the anchoring strength.
Infinite anchoring strength represents a situation where the director on an interface is fixed at the preferred alignment.
This situation is known as \emph{infinite anchoring} (sometimes also called strong anchoring).
Zero anchoring strength corresponds to a situation where the director on an interface has no preferred alignment. 
This situation is known as \emph{zero anchoring}.

Perhaps the most important effect of weak anchoring in a nematic film occurs when there is weak homeotropic anchoring on the gas--nematic interface and weak planar anchoring on the nematic--substrate interface, or vice versa.
In this situation, which is known as \emph{antagonistic anchoring}, competition between the different preferred alignments on the interfaces can introduce elastic distortion in the bulk of the nematic leading to a spatially--varying director field \cite{Lavrentovich1994,Delabre2009,Cazabat2011}, with an associated non-zero elastic energy, which can have a destabilising effect on the film \cite{LinCummings2013,Manyuhina2014}.
For situations with antagonistic anchoring, it has long been known that there exists a critical film thickness, which we term the Jenkins--Barratt--Barbero--Barberi critical thickness \cite{JenkinsBarratt,BarberoandBarberi} (often just called the Barbero--Barberi critical thickness), below which the energetically favourable state has a uniform director field in which the director aligns parallel to the preferred director alignment of the interface with the stronger anchoring. 
For film thicknesses above this critical thickness, the energetically favourable state has a director field that varies continuously across the film; this state is known as a hybrid aligned nematic (HAN) state \cite{HANDBOOK}. 

The theoretical study of nematic systems that include contact lines often avoids the consideration of antagonistic anchoring at the contact lines, by, for example, imposing infinite anchoring on the nematic--substrate interface which overrides the weak anchoring on the gas--nematic interface at the contact line (see, for example, Lam \etal \cite{LamCummings2018,LamCummings2020}), or assuming the existence of a thin precursor film on the substrate to remove the contact line entirely (see, for example, Lin \etal \cite{LinCummings2013B}).
While there have been relatively few studies of nematic contact lines, Rey \cite{Rey2003,Rey2007} considered two rather specific two-dimensional scenarios, namely either infinite planar anchoring or equal weak planar anchoring, on both interfaces.
Although neither infinite anchoring nor equal weak anchoring is likely to occur in practice, these studies highlight the possibility that anchoring breaking, \ie the process by which the preferred orientation of the nematic molecules on one of the interfaces is overridden by that on the other, occurs in a region adjacent to the contact line.
Rey \cite{Rey2003,Rey2007} also discusses the possibility of the formation of a defect, or a disclination line in his two-dimensional scenarios, located at the contact line.
At such disclination lines, a description of the nematic only in terms of the director is no longer valid and there is a high degree of elastic distortion associated with increased elastic energy \cite{SchopohlSluckin1987}.
In the present work we will assume that the energy associated with anchoring breaking in a region adjacent to the contact line is lower than the energy associated with the formation of a disclination line \cite{WaltonMcKayMottram2018} and, therefore,  that such disclination lines do not occur.

\subsection{A static ridge of liquid crystal}

Motivated by a need for increased understanding of situations involving the wetting and dewetting of nematics, in the present work we consider a two-dimensional static ridge of nematic resting on an ideal (\ie flat, rigid, perfectly smooth, and chemically homogeneous) solid substrate surrounded by a passive fluid.
In order to make comparisons with the most commonly studied experimental situation, we consider the case in which the passive fluid surrounding the nematic is an atmosphere of passive gas, although the subsequent theory and results may be readily generalised to a ridge of nematic surrounded by a static isotropic liquid.
There are many applications of liquid crystals that may benefit from an increased understanding of this situation. 
For instance, the patterning of discotic liquid crystals (discotics) into precise and controllable ridges has been demonstrated \cite{Bramble2010,Zou2018}, and this technology, together with the excellent charge-transport properties of discotics, has led to them being used as printable nanometre-scale wires for applications in electronics \cite{Sergeyev2007}.
The controlled formation of static ridges of liquid crystal also has applications in optics, particularly for creating self-organised diffraction gratings \cite{Blow2013,Brown2009}.

The nematic ridge is bounded by a gas–nematic interface and a nematic–substrate interface. The theoretical description of a nematic bounded by such interfaces has previously been considered by Jenkins and Barratt \cite{JenkinsBarratt}, who obtained general forms of the interfacial conditions and the force per unit length on a contact line, and Rey \cite{Rey2000,Rey2000b}, who obtained a general form of the nematic Young and Young--Laplace equations.
In the present work, we combine aspects of these two approaches to derive the first complete theoretical description for a static ridge of nematic, which includes the bulk elastic equation, the nematic Young equations, the nematic Young--Laplace equation, the weak anchoring conditions, and the other relevant boundary conditions. 
We provide full details of a readily accessible derivation of the governing equations in \cref{sec:model,sec:CoV,sec:GE}, which may, in principle, be generalised to include electromagnetic forces, additional contact-line effects, non-ideal substrates, or more detailed models for the nematic molecular order.

We proceed by constructing the free energy of the system as a function of both the shape of the gas--nematic interface (\ie the nematic free surface) and the director field, and then minimise the free energy using the calculus of variations. 
In order to determine the free energy of the system, we use a well-established continuum theory to consider contributions from elastic deformations of the director $\n$, the gravitational potential energy, and interface energies associated with the three interfaces (for a full account of this continuum theory of nematics, see  \cite{ISBOOK}).
We use the standard Oseen--Frank bulk elastic energy density $W$ (energy per unit volume) which depends on $\n$ and its spatial gradients \cite{ISBOOK}.
The interface energies associated with the gas--nematic and nematic--substrate interfaces will be described using the standard Rapini--Papoular  interface energy density (energy per unit area) $\omega$  which depends on $\n$ and the interface normal $\bm{\nu}$ \cite{RapiniPapoularWA}.

Although we proceed in \cref{sec:model,sec:CoV,sec:GE} by deriving the governing equations of the most commonly occurring experimental situation of the partial wetting state, $\pw$, the same governing equations also describe the complete wetting state, $\cw$, and the complete dewetting state, $\cd$.
In the  $\cw$ state, in which the nematic forms a film that completely coats the substrate, there is no gas--substrate interface and hence no contact lines. In the $\cd$ state, in which the gas--nematic interface forms a cylinder (which, because of anisotropic effects, is not necessarily circular), there is no nematic--substrate interface, and the gas--nematic interface meets the gas--substrate interface at a single contact line.
For an isotropic ridge, described briefly in \cref{sec:iso}, the classification of the equilibrium states and the transitions that occur between them are well known  and can be obtained by solving the classical isotropic Young--Laplace equation and comparing the free energies of the possible equilibrium states  \cite{Electrowetting2019,wetBOOK}.
For a nematic ridge, the free energy of the equilibrium states cannot be determined analytically; however, by comparison with the classical results for the isotropic ridge, the classification of the equilibrium states and the transitions  between them can still be obtained.
In particular, in \cref{sec:nemYE,sec:nem}, we use the nematic Young equations obtained in \cref{sec:GE} to determine the continuous and discontinuous transitions between the equilibrium states of complete wetting, partial wetting, and complete dewetting.
Previously, Rey \cite{Rey2000b} found that a general form of the general nematic Young equations allow for discontinuous transitions between partial wetting and complete wetting and between partial wetting and complete dewetting.
However, without the assumption made in the present work that anchoring breaking occurs in regions adjacent to the contact lines, an explicit description of these transitions was not possible.
Under this assumption, in \cref{sec:nemYE,sec:nem} we find not only continuous transitions analogous to those that occur in the classical case of an isotropic liquid, but also a variety of discontinuous transitions, as well as contact-angle hysteresis, and regions of parameter space in which there exist multiple partial wetting states that do not occur in the classical case.

\section{Model formulation}
\label{sec:model}

%
\begin{figure}[tp]
  \centering
  \includegraphics[width=0.85\linewidth]{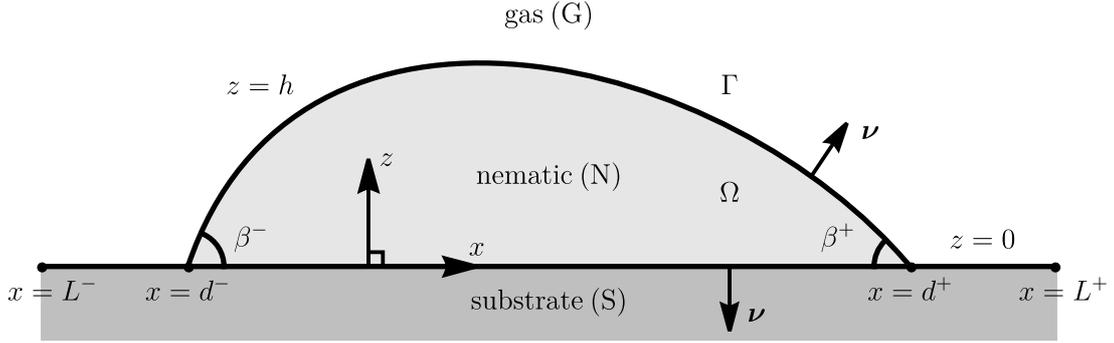} 
  \caption{A schematic of a static ridge of nematic (N) resting on an ideal solid substrate (S) at $z = 0$, $L^- \le x \le L^+$, in an atmosphere of passive gas (G), with gas--nematic interface at $z = h$ and contact lines at $x = \dpn$. The Cartesian coordinates $x$, $y$ and $z$ (where the $y$-direction is into the page), the region of nematic in the $(x,z)$-plane $\Omega$  bounded by the interface  $\Gamma$, the outward unit normals $\bm{\nu}$ and the contact angles $\betaLR$  are also indicated.
  }
\label{figRidge}
\end{figure}
As described in the previous section, we consider a static ridge of nematic (N) resting on an ideal solid substrate (S) in an atmosphere of passive gas (G), as shown in \cref{figRidge}, which also indicates the Cartesian coordinates $x$, $y$ and $z$ that we use. The region of nematic in the $(x,z)$-plane $\Omega$ is bounded by the interface $\Gamma$, which consists of the gas--nematic interface at $z=h(x)$, denoted by $\Sgn$, and the nematic--substrate interface at $z=0$, denoted by $\Sns$, and has two nematic--substrate--gas three-phase contact lines at $x=\dn$ and $x=\dpp$.
We assume that the ridge height $h$ and the position of the contact lines do not vary in the $y$-direction, so that the contact lines form two infinitely-long parallel lines in the $y$-direction and the ridge height $h$ is subject to the boundary conditions $h(\dpn)=0$. 
We also assume that the director $\n$ is confined to the $(x,z)$-plane, and hence takes the form 
\begin{align}
    \n = \cos \theta \, \unitx + \sin \theta \, \unitz,  \label{director}
\end{align}
where $\unitx$ and $\unitz$ are the Cartesian coordinate unit vectors in the $x$- and $z$-directions, respectively, and $\theta=\theta(x,z)$ is the director angle, which also does not vary in the $y$-direction.

The outward unit normals of the interfaces $\Sgn$ and $\Sns$, which we denote by $\nuGN$ and $\nuNS$, are given by
\begin{align}
    \nuGN &= -\dfrac{{h}_x}{\sqrt{1+{h}_x^2}} \unitx+\dfrac{1}{\sqrt{1+{h}_x^2}} \unitz, \label{outward1} \\
    \nuNS &= -\unitz, \label{outward2} 
\end{align}
respectively, where the subscript $x$ denotes differentiation with respect to $x$.
These two interfaces meet the gas--substrate interface, denoted by $\Sgs$, at the two contact lines $x=\dpn$. 
The left-hand and right-hand edges of the substrate are at $x=\Ln (<\dn)$ and $x=\Lp (>\dpp)$, respectively, as shown in \cref{figRidge}.
The contact angles formed between $\Sgn$ and $\Sns$ at $x=\dn$ and $x=\dpp$ are denoted by $\betaL$ and $\betaR$, respectively, and satisfy
\begin{align}
\tan \betaLR & = \mp h_x \quad \text{at} \quad x=\dpn. \label{bcA}
\end{align}
We note that there is, in general, no requirement for $h$ to be symmetric about its midpoint and, in particular, no requirement for the contact angles to be the same.

In general, we do not fix either the contact line positions or the contact angles, and allow $\dpn$ and $\betaLR$ to be unknowns. 
However, if the substrate has been treated in such a way as either to pin the contact lines or to fix the contact angles, then either $\dpn$ or $\betaLR$, respectively, are prescribed and the nematic Young equations, which will be derived shortly, are not relevant. 
The ridge has a prescribed constant cross-sectional area $A$ in the $(x,z)$-plane, so that
\begin{align}
    \iint \, \dd \Omega &= A. \label{constraint}
\end{align}

As mentioned in \cref{sec:Intro}, we include the effects of gravity.
Specifically, we assume the gravity acts in the $(x,z)$-plane but, in order to keep the setup as general as possible, do not specify its direction.

In \cref{sec:CoV} we obtain the complete theoretical description for this system using the calculus of variations assuming that the ridge height $h$ is a single-valued function of $x$. 
A necessary, but not sufficient, condition for this to be valid is that the contact angles are acute (\ie that $0 \le \betaLR \le \pi/2$). 
We have performed the corresponding derivation when the ridge height $h$ is a double-valued function of $x$. 
However, since this derivation involves either splitting the gas--nematic interface into three parts in each of which $h$ is a single-valued function of $x$ or using a different coordinate system, for simplicity of presentation, and because many of the situations described in \cref{sec:Intro} involve small contact angles, in the present work we describe the details of the derivation only when $h$ is a single-valued function of $x$.
The details of the corresponding derivation when $h$ is a double-valued function of $x$ are given by Cousins \cite{Cousinsthesis}.

\section{Constrained minimisation of the free energy}
\label{sec:CoV}

Using the calculus of variations we minimise the free energy of the system $E$ (per unit length in the $y$-direction) subject to the area constraint \cref{constraint} and the  boundary conditions $h(\dpn)=0$ to obtain the governing equations for a ridge of nematic in terms of the four unknowns $\theta(x,z)$, $h(x)$, $\dn$ and $\dpp$, and a Lagrange multiplier associated with the area constraint \cref{constraint}, the latter of which we denote by $p_0$.
The unknown contact angles $\betaLR$ are obtained from the slope of the ridge height $h_x$ using \cref{bcA}.
The free energy of the system $E$ is the sum of the bulk elastic energy of the nematic, denoted by $\Ebulk$, and the interface energies, denoted by $\Egn$, $\Ens$ and $\Egs$, for the interfaces $\Sgn$, $\Sns$ and $\Sgs$, respectively, where 
\begin{align}
   \Ebulk &= \int^{\dpp}_{\dn} \int^{h}_0  \Big( W(\theta, \theta_x, \theta_z) +\psig \Big)\, \, \dd z \, \dd x, \label{Ebulk} \\
   \Egn &= \int^{\dpp}_{\dn}  \sqrt{1+h_x^2}\,\big[ \wgn(\theta,h_x) \big]^{z=h} \, \dd x, \label{Egl} \\
   \Ens &= \int^{\dpp}_{\dn}  \big[\wns (\theta) \big]^{z=0}  \, \, \dd x, \label{Els} \\
   \Egs &=  \int^{\dn}_{\Ln} \big[ \wgs \big]^{z=0} \, \, \dd x+\int^{\Lp}_{\dpp} \big[ \wgs \big]^{z=0} \, \, \dd x. \label{Egs}
\end{align}  
In \cref{Ebulk} the bulk elastic energy density $W(\theta,\theta_x,\theta_z)$ is assumed to depend on the director angle $\theta$ and on elastic distortions of the director via the spatial derivatives of $\theta$ \cite{ISBOOK}. 
Also in \cref{Ebulk}, the gravitational potential energy density $\psig(x,z)$ is allowed to depend on one or both of the Cartesian coordinates $x$ and $z$.
In \cref{Egl,Els} the interface energy densities $\wgn(\theta,h_x)$ and $\wns(\theta)$ are assumed to be in the form of the Rapini--Papoular energy density \cite{RapiniPapoularWA}, which depends on the angle between the director \cref{director} and the outward unit normal of the interfaces, namely \cref{outward1,outward2}, respectively. 
In \cref{Egs} the interface energy density $\wgs$ takes a constant value.

We define the functional $F=F(\theta,\theta_x, \theta_z,h,h_x,\dn,\dpp)=E+\Earea$ as the sum of the free energy of the system $E$ and a term $\Earea$, corresponding to the area constraint \cref{constraint}, given by
\begin{align}   
   \Earea &= p_0 \times \left( A- \int^{\dpp}_{\dn} \int^{h}_0  \, \dd z \, \dd x\right), \label{Earea}
\end{align}
so that the functional $F$ is given by 
\begin{align}
    F=\Ebulk+\Egn+\Ens+\Egs+\Earea. \label{energy} 
\end{align}
We now consider the variation of $F$, given by \cref{energy} with \cref{Ebulk,Egl,Els,Egs,Earea}, with respect to small variations of the variables $\theta$, $h$, $\dn$ and $\dpp$ of the form
\begin{align}
    \theta \to \theta + \delt, \quad h \to h + \delh, \quad  \dn \to \dn + \deldn, \quad \dpp \to \dpp + \deldp. \label{var}
\end{align}
There are no constraints on the director angle $\theta$, and therefore there are no constraints on its variation $\delt$.
There is, however, a constraint on the ridge height $h$ because of the boundary conditions $h(\dpn)=0$, so that the variation of the ridge height $\delh$ at the contact lines satisfies
\begin{align}
    \delh &= - {h}_x \deldn=-\tan\betaL \, \deldn \quad \text{at} \quad x=\dn, \label{endpoint1} \\
    \delh &= - {h}_x \deldp= +\tan\betaR \, \deldp \quad \text{at} \quad x=\dpp. \label{endpoint2}
\end{align}

The variation of the functional $F$, denoted by $\delta F$, is given by
\begin{align}
    \delta F &= F\left(\theta+\delt,(\theta+\delt)_x, (\theta+\delt)_z,h+\delh,(h+\delh)_x,\dn+\deldn,\dpp+\deldp)\right. \nonumber \\ 
    &\qquad \qquad \left.-F(\theta,\theta_x, \theta_z,h,h_x,\dn,\dpp\right). \label{delF1}
\end{align}
We now consider the variation of each term in \cref{energy} in turn, and neglect terms in \cref{delF1} that are quadratic in the variations $\delt$, $\delh$, $\deldn$ and $\deldp$. 

For the bulk elastic energy $\Ebulk$, given by \cref{Ebulk}, $\delta \Ebulk$ is given by 
\begin{align*}
  \delta \Ebulk &= \int^{\dpp}_{\dn} \int^{h}_0  \delt \pdv{W}{\theta} + {\delt}_x \pdv{W}{\tx}+ {\delt}_z \pdv{W}{\tz}\, \, \dd z \, \dd x  +   \int^{\dpp}_{\dn} \delh \left[ W +\psig \right]^{z=h} \, \, \dd x \\
  & \quad \quad - \deldn\Bigg[ \int^{h}_{0}  (W+\psig) \, \, \dd z \Bigg]^{x=\dn}+  \deldp\Bigg[ \int^{h}_{0}  (W+\psig) \, \, \dd z \Bigg]^{x=\dpp}. \numberthis \label{bulkstep1}
\end{align*}
Since $h(\dpn)=0$, the last two terms in \cref{bulkstep1} are identically zero.
The terms in \cref{bulkstep1} containing derivatives of $\delt$, namely ${\delt}_x$ and ${\delt}_z$, are transformed into terms involving $\delt$ by using the divergence theorem, namely  
\begin{align}
    \int \int {\delta_{\theta}}_\alpha \pdv{W}{\theta_\alpha} \, \dd \Omega &= \oint_\Gamma \delt \pdv{W}{\theta_\alpha} \bm{\hat{\alpha}} \cdot \bm{\nu} \, \dd \Gamma -  \int \int \delt \pdv{\alpha} \left(\pdv{W}{\theta_\alpha}\right) \, \dd \Omega, \label{divergence}
\end{align} 
where $\alpha=x$ or $\alpha=z$.
The line integral along $\Gamma$ in \cref{divergence} is composed of a component along $\Sgn$ from $x=\dpp$ to $x=\dn$ on $z=h$ with $\dd \Gamma = -\sqrt{1+{h}_x^2} \, \dd x$ and outward unit normal \cref{outward2}, and a component along $\Sns$ at $z=0$ from $x=\dn$ to $x=\dpp$ with $\dd \Gamma= \dd x$ and outward unit normal \cref{outward1}, and is given explicitly by 
\begin{align*}
\oint_\Gamma \delt \pdv{W}{\theta_\alpha} \bm{\hat{\alpha}} \cdot \bm{\nu} \, \dd \Gamma &=  \int^{\dn}_{\dpp} \Bigg[ \delt \pdv{W}{\theta_\alpha} \Bigg]^{z=h} \bm{\hat{\alpha}} \cdot (h_x \unitz-\unitz) \, \, \dd x- \int^{\dpp}_{\dn} \Bigg[ \delt \pdv{W}{\theta_\alpha} \Bigg]^{z=0} \bm{\hat{\alpha}} \cdot \unitz \, \, \dd x  \numberthis  \label{bulkstep2}.
\end{align*}
Equations \cref{bulkstep1}--\cref{bulkstep2} can be combined and rearranged to express the variation $\delta E_{\rm bulk}$ as
\begin{align*}
   \delta E_{\rm bulk} &= \int^{\dpp}_{\dn} \int^{h}_0 \delt \left( \pdv{W}{\theta} - \pdv{x} \left( \pdv{W}{\tx}\right)- \pdv{z} \left( \pdv{W}{\tz}\right) \right) \, \dd z \, \dd x  \\
    & \quad \quad +\int^{\dpp}_{\dn}  \delh  \left[  W+\psig \right]^{z=h} \, \, \dd x-\int_{\dn}^{\dpp} \left[ \delt h_x \pdv{W}{\tx} \right]^{z=h} \, \, \dd x \numberthis \label{delbulk} \\
    & \quad \quad+ \int_{\dn}^{\dpp} \left[ \delt \pdv{W}{\tz} \right]^{z=h} \, \, \dd x- \int_{\dn}^{\dpp} \left[ \delt \pdv{W}{\tz} \right]^{z=0}\dd x.  
\end{align*}

For the gas--nematic interface energy $\Egn$, given by \cref{Egl}, carrying out integration by parts on the terms involving ${\delh}_x$ shows that $\delta \Egn$ is given by 
\begin{align*}
    \delta \Egn &=  \int^{\dpp}_{\dn} \left[ \delt\sqrt{1+{h}_x^2} \pdv{\wgn}{\theta} + \delh \left(  \sqrt{1+{h}_x^2} \pdv{\wgn}{\theta} \pdv{\theta}{z} - \pdv{x}\Big[ \pdv{{h}_x} \big( \sqrt{1+{h}_x^2} \, \, \wgn  \big) \Big] \right) \right]^{z=h} \, \dd x \\
    & \quad \quad  - \deldn \left[\sqrt{1+{h}_x^2} \, \wgn \right]^{x=\dn}+\deldp \left[\sqrt{1+{h}_x^2} \, \wgn \right]^{x=\dpp} \numberthis \label{GLstep1} \\
   & \quad \quad -\left[  \delh \pdv{{h}_x} \left( \sqrt{1+{h}_x^2} \, \, \wgn \right) \right]^{x=\dn}+ \left[  \delh \pdv{{h}_x} \left( \sqrt{1+{h}_x^2} \, \, \wgn \right) \right]^{x=\dpp}. 
\end{align*}
Substituting for the variation of the ridge height $\delh$ at the contact lines, given by \cref{endpoint1,endpoint2}, then yields
\begin{align*}
    \delta \Egn &= \int^{\dpp}_{\dn} \Bigg[ \delt\sqrt{1+{h}_x^2} \pdv{\wgn}{\theta} + \delh \left(  \sqrt{1+{h}_x^2} \pdv{\wgn}{\theta} \pdv{\theta}{z} -\pdv{x}\Big[ \pdv{{h}_x} \big( \sqrt{1+{h}_x^2} \, \, \wgn  \big) \Big] \right) \Bigg]^{z=h} \, \dd x \\
    &  \quad \quad - \deldn \left[\sqrt{1+{h}_x^2} \, \wgn -  {h}_x  \pdv{{h}_x} \left(\sqrt{1+{h}_x^2} \, \wgn \right) \right]^{x=\dn}\label{delGL} \numberthis \\
    &  \quad \quad +\deldp \left[\sqrt{1+{h}_x^2} \, \wgn -  {h}_x  \pdv{{h}_x} \left(\sqrt{1+{h}_x^2} \, \wgn \right) \right]^{x=\dpp} .
\end{align*}

For the nematic--substrate interface energy $\Ens$, given by \cref{Els},  $\delta \Ens$ is given by
\begin{align}
\delta \Ens  &= \int^{\dpp}_{\dn} \left[\delt \pdv{\wns}{\theta} \right]^{z=0} \, \, \dd x -\deldn\left[ \wns\right]^{x=\dn}+\deldp\left[ \wns\right]^{x=\dpp}. \label{delLS} 
\end{align}

For the gas--substrate interface energy $\Egs$, given by \cref{Egs}, $\delta \Egs$ is given by
\begin{align}
  \delta E_{\rm GS}  &= \deldn\left[ \wgs \right]^{x=\dn}- \deldp\left[ \wgs \right]^{x=\dpp}. \label{delGS}
\end{align}

Finally, for the area constraint term $\Earea$, given by \cref{Earea}, using the boundary conditions $h(\dpn)=0$ shows that $\delta \Earea$ is given by
\begin{align}
\delta \Earea &= -\int^{\dpp}_{\dn} \delh\, p_0   \, \, \dd x. \label{delA}
\end{align}

The variation of $F$ is obtained by adding the terms from each of the individual variations, given by \cref{delbulk} and \cref{delGL}--\cref{delA}, so that 
\begin{align}
    \delta F =\delta \Ebulk+\delta \Egn+\delta \Ens+ \delta \Egs+\delta \Earea. \label{delF}
\end{align}
Since we seek extrema of the free energy $E$ for which $\delta F=0$, and the variations $\delt$, $\left[\delt\right]^{z=0}$, $\left[\delt\right]^{z=h}$, $\delh$, $\deldn$ and $\deldp$  are independent and arbitrary, their coefficients in $\delta F$, given by \cref{delF}, must be zero. 
Together with the area constraint \cref{constraint} and the boundary conditions $h(\dpn)=0$, the coefficients of each variation yield the governing equations for a nematic ridge, as described in the next section.

\section{Governing equations for a nematic ridge}
\label{sec:GE}

Each of the six governing equations derived from setting the coefficients of $\delt$,  $\left[\delt\right]^{z=0}$, $\left[\delt\right]^{z=h}$, $\delh$, $\deldn$ and $\deldp$  in \cref{delF} to zero has a distinct physical interpretation, namely the balance of elastic torque within the bulk of the nematic, the balance-of-couple conditions on the gas--nematic and nematic--substrate interfaces, the balance-of-stress condition on the gas--nematic interface, and the balance-of-stress conditions at the contact lines, respectively.
These equations are summarised below.

The balance of elastic torque within the bulk of the nematic, \ie the Euler--Lagrange equation, for the elastic free energy density $W$, is
\begin{align}
    \pdv{W}{\theta}-\pdv{x}\left(\pdv{W}{\theta_x} \right)-\pdv{z}\left(\pdv{W}{\theta_z} \right)&=0. \label{EqBulk}
\end{align}

The balance-of-couple conditions on the gas--nematic interface and the nematic--substrate interface, namely the weak anchoring conditions \cite{ISBOOK,HANDBOOK_LCMODELLING}, are given by
\begin{align}
    \pdv{W}{\theta_z}-{h}_x \, \pdv{W}{\theta_x} + \sqrt{1+{h}_x^2} \, \pdv{\wgn}{\theta}&=0 \quad \text{on} \quad z=h \label{EqGL} 
\end{align}
and
\begin{align}
    -\pdv{W}{\theta_z}+\pdv{\wns}{\theta}&=0 \quad \text{on} \quad z=0, \label{EqLS}
\end{align}
respectively. 

The balance-of-stress condition on the gas--nematic interface is given by
\begin{align}
    W+\psig-p_0+ \sqrt{1+{h}_x^2} \, \pdv{\wgn}{\theta} \pdv{\theta}{z} - \pdv{x}\left( \pdv{{h}_x} \left( \sqrt{1+{h}_x^2} \, \, \wgn  \right) \right)&=0\quad \text{on} \quad z=h. \label{EqYL} 
\end{align}
To distinguish equation \cref{EqYL} from the classical isotropic Young--Laplace equation \cite{wetBOOK}, henceforth it is referred to as the \emph{nematic Young--Laplace equation}.

The balance-of-stress conditions at the contact lines are given by
\begin{align}
  \wns - \wgs + \sqrt{1+{h}_x^2} \, \wgn -{h}_x \pdv{{h}_x} \left(\sqrt{1+{h}_x^2} \, \wgn \right)&=0  \quad \text{at} \quad x=\dpn. \label{EqYE}
\end{align}
To distinguish equations \cref{EqYE} from the classical isotropic Young equations \cite{wetBOOK}, henceforth they are referred to as the \emph{nematic Young equations}.

Once explicit forms of the energy densities $W$, $\psig$, $\wgn$, $\wns$ and $\wgs$ have been prescribed, the balance of elastic torque within the bulk of the nematic \cref{EqBulk}, the three interface conditions \cref{EqLS,EqGL,EqYL}, the two nematic Young equations \cref{EqYE}, the area constraint \cref{constraint}, and the two boundary conditions $h(\dpn)=0$ provide the complete theoretical description for a static ridge of nematic in terms of the five unknowns $\theta(x,z)$, $h(x)$, $\dn$, $\dpp$ and $p_0$.

\subsection{The bulk elastic energy density and the interface energy densities}

As mentioned in \cref{sec:Intro}, for the bulk elastic energy density $W$ we use the standard Oseen--Frank bulk elastic energy density \cite{ISBOOK} for which
\begin{align}
    W &=  \dfrac{1}{2} K_1 (\nabla \cdot \n)^2+\dfrac{1}{2} K_2 (\n \cdot \nabla \times \n)^2+ \dfrac{1}{2} K_3 (\n \times \nabla \times \n)^2 \nonumber \\
    & \quad \quad  +\dfrac{1}{2} (K_2+K_4) \nabla \cdot \big[(\n \cdot \nabla) \n - ( \nabla \cdot \n)\n\big],  \label{energybulk0}
\end{align}
where $K_1$, $K_2$, $K_3$, and the combination $K_2+K_4$ are called the splay, twist, bend, and saddle-splay elastic constants, respectively, and $\nabla=(\partial/\partial x,\partial/\partial y,\partial/\partial z)$. 
Substituting \cref{director} into \cref{energybulk0} yields
\begin{align}
    W(\theta,\theta_x,\theta_z)&=\dfrac{K_1}{2} \left(\theta_z \cos\theta-\theta_x\sin\theta \right)^2+\dfrac{K_3}{2}\left(\theta_x\cos\theta+\theta_z\sin\theta\right)^2, \label{energybulk}
\end{align}
which depends only on the splay and bend elastic deformations. 
Although we use the full Oseen--Frank energy density \cref{energybulk}, we note that the commonly used one-constant approximation of the elastic constants \cite{ISBOOK} can be implemented in \cref{energybulk} by setting $K=K_1=K_3$, in which case $W(\theta_x,\theta_z)= K(\theta_x^2 +\theta_z^2)/2$.

As also mentioned in \cref{sec:Intro}, for $\wgn$ and $\wns$ we use the standard Rapini--Papoular form \cite{RapiniPapoularWA} for which
\begin{align}
\wgn &= \ggn + \dfrac{\cgn}{4} \left( 1 - 2(\nuGN \cdot \n)^2 \right),  \label{energygl0} \\
\wns &= \gns + \dfrac{\cns}{4} \left( 1 - 2(\nuNS \cdot \n)^2 \right),  \label{energyls0}
\end{align}
where $\cgn$ and $\ggn$ are the anchoring strength and isotropic interfacial tension for the gas--nematic (GN) interface, respectively, and  $\cns$ and $\gns$ are the anchoring strength and isotropic interfacial tension for the nematic--substrate (NS) interface, respectively.
The Rapini--Papoular form ensures that the interface energy densities $\wgn$ and $\wns$ are at a minimum when $\n$ and $\bm{\nu}$ are parallel for $\cgn>0$ and $\cns>0$, respectively, and at a minimum when $\n$ and $\bm{\nu}$ are perpendicular for $\cgn<0$ and $\cns<0$, respectively. 
Therefore weak homeotropic anchoring occurs on the gas--nematic interface when $\cgn > 0$ and on the nematic--substrate interface when $\cns > 0$, and weak planar anchoring  occurs on the gas--nematic interface when $\cgn < 0$ and on the nematic--substrate interface when $\cns < 0$.
Substituting \cref{outward1,outward2} into \cref{energygl0,energyls0} yields
\begin{align}
 \wgn(\theta, {h}_x) &= \ggn+\dfrac{\cgn}{4} \left[\dfrac{1-{h}_x^2}{1+{h}_x^2} \cos 2 \theta+\dfrac{2 {h}_x}{1+{h}_x^2} \sin 2 \theta \right],  \label{energygl} \\
 \wns(\theta) &= \gns+\dfrac{\cns}{4} \cos 2 \theta. \label{energyls} 
\end{align}

Experimental techniques for the measurement of $\cns$ are well established \cite{AnchoringMeasurement1,AnchoringMeasurement2,Yokoyama1988}, and values in the range $|\cns| = 10^{-6}$--$10^{-3}\,$N$\,$m$^{-1}$ have been reported for a variety of nematic materials and substrates with planar or homeotropic anchoring \cite{AnchoringMeasurement1,AnchoringMeasurement2,SONINBOOK}.
Measurements of $\cgn$ are less common \cite{SONINBOOK}; however, the reported values of $\cgn>10^{-5}\,{\rm N m}^{-1}$ between air and the nematic mixture ZLI 2860 \cite{FreeSurfaceAnchoring} and of $\cgn>10^{-4}\,{\rm N m}^{-1}$ between air and the nematic p-methoxy-benzylidene-p-n-butyl aniline (MBBA) \cite{Chiarelli1983} suggest that $\cgn$ and $\cns$ can be of comparable magnitude.
In standard low-molecular-mass nematics, the isotropic interfacial tensions (\ie $\ggn$ and $\gns$) are typically much larger than the magnitudes of the anchoring strengths (\ie $|\cgn|$ and $|\cns|$) \cite{SONINBOOK}.
For example, the isotropic interfacial tension of an interface between air and the nematic 4-Cyano-4'-pentylbiphenyl (5CB) has been measured as $\ggn=4.0\times10^{-2}\,{\rm N\, m}^{-1}$, and the isotropic interfacial tension of an interface between the substrate poly(methyl methacrylate) (PMMA) and 5CB has been measured as $\gns=4.051\times10^{-2}\,{\rm N\,m}^{-1}$ \cite{Dhara2020}.

The gas--substrate interface has constant energy density
\begin{align}
    \wgs = \ggs, \label{energygs} 
\end{align}
where $\ggs$ is the isotropic interfacial tension of the gas--substrate interface.

\subsection{Governing equations using the Oseen--Frank bulk elastic energy density and the Rapini--Papoular interface energy densities}
\label{sec:GE2}

Using \cref{energybulk} in \cref{EqBulk} yields the balance of elastic torque within the bulk of the nematic,
\begin{align}
    &(K_1 \sin^2\theta +K_3 \cos^2 \theta)\theta_{xx} + (K_1 \cos^2\theta +K_3 \sin^2 \theta)\theta_{zz}
\label{EqBulkF}\\
    & \quad \quad+(K_3-K_1)\Big[\left(\theta_z \cos\theta-\theta_x\sin\theta \right)\left(\theta_x \cos\theta+\theta_z \sin\theta \right)+ \theta_{xz}\sin2\theta\Big]=0. \nonumber
\end{align}
Using \cref{energybulk,energygl} in \cref{EqGL} yields the balance-of-couple condition on the gas--nematic interface,
\begin{align}
    &\left(K_1\cos^2 \theta +K_3\sin^2\theta\right)\theta_z+\dfrac{1}{2}(K_3-K_1)(\theta_x-h_x\theta_z)  \sin2\theta -\left(K_1\sin^2 \theta +K_3\cos^2\theta\right)h_x\theta_x \nonumber \\
    & \quad \quad+\dfrac{\cgn}{2\sqrt{1+{h}_x^2}} \left[({h}_x^2-1) \sin 2\theta +2 {h}_x \cos 2\theta\right]=0 \quad \text{on} \quad z=h, \label{EqGLF} 
\end{align}
while using \cref{energyls} in \cref{EqLS} yields the balance-of-couple condition on the nematic--substrate interface,
\begin{align}
    -\left(K_1\cos^2 \theta +K_3\sin^2\theta\right)\theta_z-\dfrac{1}{2}(K_3-K_1)\theta_x \sin2\theta-\dfrac{\cns}{2} \sin 2\theta=0 \quad \text{on} \quad z=0.\label{EqLSF} 
\end{align}
Using \cref{energybulk,energygl} in \cref{EqYL} yields the nematic Young--Laplace equation,
\begin{align}
   &p_0-W-\psig+\ggn \dfrac{{h}_{xx}}{\left( 1+{h}_x^2\right)^{3/2}}+\dfrac{\cgn}{4\left( 1+{h}_x^2\right)^{5/2}} \Bigg[ 3 {h}_{xx} \Big[({h}_x^2-1)\cos 2 \theta-2 {h}_x \sin2\theta  \Big] \label{EqYLF} \nonumber \\
   & \quad \quad   + (1+{h}_x^2) \bigg(4\cos2\theta\Big[\theta_x-h_x(1+{h}_x^2)\theta_z\Big]+2\sin2 \theta \Big[(1-{h}_x^4)\theta_z+{h}_x(3+{h}_x^2) \theta_x\Big] \bigg) \Bigg]=0 \nonumber\\
   & \hspace{9.55cm} \text{on} \quad z={h}.
\end{align}
In order to express the nematic Young equations \cref{EqYE} in terms of the contact angles $\betaLR$, we use the relations \cref{bcA}.
Then, using \cref{energygl,energyls,energygs} in \cref{EqYE} yields
\begin{align}
    \ggs  - \gns - \ggn \cos \betaL &= \dfrac{\cns}{4} \cos 2\theta + \dfrac{\cgn}{4}\left[ \cos2(\theta-\betaL) \cos \betaL -2 \sin2(\theta - \betaL) \sin \betaL \right]\nonumber \\
   & \qquad \qquad \qquad \qquad \qquad \qquad \qquad \qquad \qquad \qquad \text{at} \quad x=\dn, \label{EqYEF1} \\
   \ggs  - \gns - \ggn \cos \betaR &= \dfrac{\cns}{4} \cos 2\theta + \dfrac{\cgn}{4}\left[ \cos2(\theta+\betaR) \cos \betaR -2 \sin2(\theta +\betaR) \sin \betaR \right] \nonumber \\
   & \qquad \qquad \qquad \qquad \qquad \qquad \qquad \qquad \qquad \qquad \text{at} \quad x=\dpp. \label{EqYEF2} 
\end{align}
The terms on the left-hand sides of \cref{EqYEF1,EqYEF2} appear in the classical isotropic Young equations, while the terms on the right-hand sides are due to the anisotropic nature of the nematic and arise from the weak anchoring on the gas--nematic interface and on the nematic--substrate interface, respectively. 
In particular, the classical isotropic Young equations are recovered from the nematic Young equations \cref{EqYEF1,EqYEF2} by setting $\cns=\cgn=0$.

We note that, as previously mentioned, while the nematic Young equations \cref{EqYEF1,EqYEF2} were derived assuming the ridge height $h$ is a single-valued function of $x$, they also hold when the ridge height $h$ is a double-valued function of $x$ \cite{Cousinsthesis}.

\subsection{The equilibrium states of complete wetting and complete dewetting}
\label{sec:CDW:CW}

The governing equations derived thus far in the present manuscript describe the partial wetting state, $\pw$.
As mentioned in \cref{sec:Intro}, these equations can also be used to describe the equilibrium states of complete wetting, $\cw$, and of complete dewetting,  $\cd$. In the $\cw$ state, in which the nematic forms a film that completely coats the substrate, the nematic Young equations \cref{EqYEF1,EqYEF2} and the boundary conditions $h(\dpn)=0$ are not relevant.  
The behaviour of the director and gas--nematic interface for nematic films has been studied previously (see, for example, Sonin \cite{SONINBOOK} and Manyuhina \cite{Manyuhina2014}).
Similarly, for the $\cd$ state, in which the gas--nematic interface forms a cylinder, the nematic Young equations \cref{EqYEF1,EqYEF2}, the boundary conditions $h(\dpn)=0$, and the balance-of-couple condition on the nematic--solid interface \cref{EqLSF} are not relevant. 
In the special case in which the gas--nematic interface is a circular cylinder, the possible director configurations are the same as those in the case of a nematic confined within a circular capillary and have been extensively studied (see, for example, Kleman \cite{KlemanBook}). 
The limiting cases $\betaLR=0$ and $\betaLR=\pi$ correspond to the $\cw$ and the $\cd$ states, respectively.

As we will show in what follows, using just the nematic Young equations \cref{EqYEF1,EqYEF2} under the assumption that anchoring breaking occurs in regions adjacent to the contact lines, we can determine the continuous and discontinuous transitions that occur between the equilibrium states of complete wetting, partial wetting, and complete dewetting.
We first briefly review the behaviour of an isotropic ridge in \cref{sec:iso}, before analysing the corresponding behaviour of a nematic ridge in \cref{sec:nemYE,sec:nem}.

\section{The equilibrium states and transitions of an isotropic ridge}
\label{sec:iso}

For a static ridge of isotropic liquid resting on an ideal solid substrate in an atmosphere of passive gas, a much simpler version of the derivation presented in \cref{sec:model,sec:CoV,sec:GE}, or a simple horizontal force balance at the contact lines, shows that the well-known classical isotropic Young equations \cite{wetBOOK} are given, in the present notation, by
\begin{align}
\ggs-\gis-\ggi \cos \betaLR=0 \quad \text{at} \quad x=\dpn, \label{isoYE1} 
\end{align}
where $\ggi$ and $\gis$ denote the isotropic interfacial tensions of the gas--isotropic liquid and isotropic liquid--substrate interfaces, respectively.
In particular, \cref{isoYE1} shows that in this case the left-hand and right-hand contact angles are always the same, \ie $\betaL=\betaR=\beta$, say.
The classical isotropic Young equations \cref{isoYE1} can be written in terms of a single non-dimensional parameter, namely the classical isotropic spreading parameter $\Si$, which is defined by
\begin{align}
\Si=\dfrac{\ggs - \gis}{\ggi} - 1, \label{Si} 
\end{align}
as
\begin{align}
    \Si+1-\cos \beta=0. \label{isoYE}
\end{align}
Specifically, \cref{isoYE} shows that the $\pw$ state exists only when $-2 \le \Si \le 0$ and that the contact angle is then given by $\beta = \cos^{-1} \left(\Si+1\right)$.

\begin{figure}[tp]
\begin{center}
 \includegraphics[width=7cm]{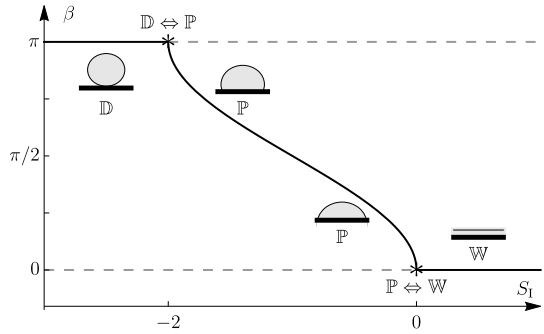} 
\end{center}
\caption{
Summary of the solution for the contact angle $\beta$ as a function of the isotropic spreading parameter $\Si$. The transition points are denoted by stars $(\Ast)$. The solid
line denote the local minimum energy state and the dashed lines denote the local maximum energy states. Sketches of the minimum energy state are also shown.}
\label{bifurcationISO}
\end{figure}

As mentioned in Sections \ref{sec:Intro} and \ref{sec:CDW:CW}, the equilibrium state can also be the $\cw$ state, which corresponds to $\beta=0$, or the $\cd$ state, which corresponds to $\beta=\pi$, for both of which equation \cref{isoYE} is not relevant.
The classification of the $\cw$, $\pw$ and $\cd$ states can be obtained by solving the classical isotropic Young--Laplace equation and expressing the minimum energy state in terms of  $\Si$ \cite{Electrowetting2019,wetBOOK}.
In particular, the minimum energy state is the $\cw$ state for $\Si>0$, the $\pw$ state for $-2 \le \Si\le 0$, and the $\cd$ state for $\Si <-2$.
The contact angle $\beta$ of the minimum energy state of an isotropic ridge is plotted as a function of $\Si$ in \cref{bifurcationISO}.
%

We define the values of $\Si$ at which there is a change in the number of possible equilibrium states as \emph{transition points}.
At these points, a transition occurs as $\Si$ increases or decreases if the previous minimum energy state ceases to exist or a new minimum energy state comes into existence. In particular, as \cref{bifurcationISO} shows, for an isotropic ridge there is a change in the number of equilibrium states at the transition points $\Si=-2$ and $\Si=0$ which leads to continuous transitions to a new minimum energy state as $\Si$ increases or decreases.
For consistency with the notation used  in \cref{sec:nem}, we denote a continuous transition between two equilibrium states for both increasing and decreasing $\Si$ with a double arrow $(\Leftrightarrow)$. 
At $\Si=-2$ there is a continuous transition from complete dewetting to partial wetting or vice versa, which is denoted by $\cd \Leftrightarrow \pw$. 
Similarly, at $\Si=0$ there is a continuous transition from complete wetting to partial wetting or vice versa, which is denoted by $\pw \Leftrightarrow \cw$.

We also note that the behaviour of the contact angle for an isotropic ridge is non-hysteretic. 
The well-known phenomenon of isotropic contact-angle hysteresis occurs only in isotropic systems with \emph{non-ideal} substrates \cite{wetBOOK}, and therefore does not occur for the isotropic ridge on an ideal substrate discussed in this section.

\section{The nematic Young equations} 
\label{sec:nemYE}

As for the isotropic ridge discussed in the previous section, for the nematic ridge considered in the present work we can use the nematic Young equations \cref{EqYEF1,EqYEF2} to determine the continuous and discontinuous transitions that occur between the equilibrium states of complete wetting, partial wetting, and complete dewetting.
At first sight, determining these transitions would appear to involve solving the governing equations for $\theta$ in the bulk of the nematic ridge, which would, in turn, involve solving for the ridge height $h$, the contact line positions $x=\dpn$, and the Lagrange multiplier $p_0$.
However, under the assumption that anchoring breaking occurs in regions adjacent to the contact lines, we can determine these continuous and discontinuous transitions from just the nematic Young equations \cref{EqYEF1,EqYEF2}.

\subsection{The director orientation at the contact lines}
\label{sec:thetaCL}

At the contact lines the preferred director orientations on the gas--nematic and the nematic--substrate interfaces are, in general, different. 
Even when the anchoring is non-antagonistic (\ie when either planar or homeotropic anchoring is preferred on both interfaces), since the preferred director orientation of both interfaces is measured relative to that interface, and the two interfaces meet at the non-zero contact angles $\betaLR$, the orientations are, in general, not the same.
Hence the director cannot, in general, align with the preferred orientations of both interfaces.
In such a situation there are three possibilities for the director orientation at the contact lines: 
(i) the contact angles are such that the preferred orientations on the two interfaces coincide exactly; 
(ii) there may be defects (disclination lines in this two-dimensional case) at one or both of the contact lines;
(iii) the weak anchoring on both interfaces allows anchoring breaking to occur in regions adjacent to the contact lines and the director(s) on one or both of the interfaces deviates from the preferred alignment(s) and attains the same orientation on both interfaces.


Case (i) is a very special situation in which the contact angles are such that the preferred director orientations on the two interfaces coincide exactly at the contact lines. For instance, when the preferred orientations on the two interfaces are antagonistic, the contact angles must be exactly $\betaLR=\pi/2$ to allow the director to be tangent to one interface and perpendicular to the other. Since this type of special case is highly unlikely to occur in practice, we do not consider it any further in the present work.


As discussed in \cref{sec:Intro}, case (ii) has been considered in \cite{Rey2003} in which infinite planar anchoring was assumed on the gas--nematic and nematic--substrate interfaces.
In this case, since the infinitely-strong anchoring cannot be broken, the director must adopt a splayed configuration (for a full account of splayed director configurations, see \cite{ISBOOK}) in a region adjacent to the contact line, with a disclination line located at the contact line \cite{Rey2003}.
For the finite anchoring strengths considered in the present work, we assume that the energy associated with anchoring breaking is less than the energy associated with the formation of a disclination line, and therefore that such disclination lines do not occur.


Having ruled out cases (i) and (ii), we are left with case (iii).
In this case the weak anchoring on the interfaces allows anchoring breaking to occur in regions adjacent to the contact lines so that the director(s) on one or both of the interfaces deviates from the preferred alignment(s) and attains the same orientation on both interfaces.

As discussed in \cref{sec:Intro}, for nematic films with antagonistic anchoring, when the film thickness is less than the Jenkins--Barratt--Barbero--Barberi critical thickness the energetically favourable state has a uniform director field in which the director aligns parallel to the preferred director alignment of the interface with the stronger anchoring.
For a nematic ridge,  close to the contact lines, where the ridge height approaches zero and hence the separation between the gas--nematic and nematic--substrate interfaces is always less than the critical thickness, anchoring breaking occurs and the director aligns parallel to the preferred alignment of the interface with the stronger anchoring.
Specifically, if the nematic--substrate interface has the stronger anchoring (\ie if $|\cns|>|\cgn|$) then the director at the contact lines aligns parallel to the nematic--substrate interface with $\theta=0$ at $x=\dpn$ in the case of planar anchoring corresponding to $\cns<0$ or perpendicular to the nematic--substrate interface with $\theta=\pi/2$ at $x=\dpn$ in the case of homeotropic anchoring corresponding to $\cns>0$; we term both of these scenarios ``nematic--substrate (NS) dominant anchoring''.
Correspondingly, if the gas--nematic interface has the stronger anchoring (\ie if $|\cgn|>|\cns|$) then the director at the contact lines aligns parallel to the gas--nematic interface with $\theta=\betaLR$ at $x=\dpn$ in the case of planar anchoring corresponding to $\cgn<0$ or perpendicular to the gas--nematic interface with $\theta=\betaLR+\pi/2$ at $x=\dpn$ in the case of homeotropic anchoring corresponding to $\cgn>0$; we term both of these situations ``gas--nematic (GN) dominant anchoring''.

There are two special situations in which anchoring breaking cannot occur as described above because the interfaces have either equal anchoring strengths ($\cns=\cgn$) or equal and opposite anchoring strengths ($\cns=-\cgn$).
In both of these situations, anchoring breaking occurs on both interfaces and the director orientation adopts the average of the preferred orientations \cite{Cousinsthesis,Rey2003}. 
In particular, when the anchoring strengths of the interfaces are equal and planar anchoring is preferred, the director angles are $\theta=\betaLR/2$ at $x=\dpn$, as discussed by Rey \cite{Rey2003},
and when the anchoring strengths of the interfaces are equal and homeotropic anchoring is preferred, the director angles are $\theta=\betaLR/2+\pi/2$ at $x=\dpn$.
When the anchoring strengths of the interfaces are equal and opposite, the director angles are $\theta=\betaLR/2+\pi/4$ or $\theta=\betaLR/2-\pi/4$ at $x=\dpn$.

Since for an ideal substrate the material properties of the substrate are the same at both contact lines, anchoring breaking must occur in the same way, and hence the director angles at the two contact lines must be the same.
However, as we will show below, in some situations the nematic Young equations \cref{EqYEF1,EqYEF2} allow for more than one possible contact angle for the same parameter values, and so the contact angles $\betaLR$ do not, in general, have to be the same and so the ridge can be asymmetric.
Moreover, the contact angles $\betaLR$ could be different if the substrate is non-ideal and the material properties of the substrate are different at the two contact lines  (for example, if the substrate was manufactured so that the values of $\cns$ at $x=\dpn$ were different, or if gradients in the temperature of the gas or adsorption of a surfactant from the gas lead to different values of $\cgn$ at $x=\dpn$ \cite{McCamley2009}).
Without loss of generality, for the remainder of the present work, we consider only the left-hand contact line, which is described by the nematic Young equation \cref{EqYEF1}, and write $\betaL=\beta$ for simplicity.
The corresponding results for the right-hand contact line can be obtained in the same way.

\subsection{Nematic spreading parameters}

For NS-dominant anchoring (for which either $\theta = 0$ or $\theta=\pi/2$) the nematic Young equation \cref{EqYEF1} reduces to a cubic equation for $\cos \beta$, namely either
\begin{align}
   \ggs - \left(\gns+\tfrac{1}{4}\cns\right)- \left(\ggn+\tfrac{1}{4}\cgn\right)\cos \beta  = -\tfrac{1}{2}\cgn \cos \beta \left(\cos^2 \beta-1\right) \label{LSdomA} 
\end{align}
when $\theta=0$ or
\begin{align}   
   \ggs - \left(\gns-\tfrac{1}{4}\cns\right)- \left(\ggn-\tfrac{1}{4}\cgn\right)\cos \beta   = \tfrac{1}{2}\cgn \cos \beta \left(\cos^2 \beta-1\right) \label{LSdomB}
\end{align}
when $\theta=\pi/2$.
On the other hand, for GN-dominant anchoring (for which either $\theta = \beta$ or $\theta=\beta + \pi/2$) the nematic Young equation \cref{EqYEF1} reduces to a quadratic equation for $\cos \beta$, namely either
\begin{align}
   \ggs - \left(\gns+\tfrac{1}{4}\cns\right)- \left(\ggn+\tfrac{1}{4}\cgn\right)\cos \beta  = \tfrac{1}{2}\cns  \left(\cos^2 \beta-1\right) \label{GLdomA}
\end{align}
when $\theta=\beta$ or
\begin{align}  
   \ggs - \left(\gns-\tfrac{1}{4}\cns\right)- \left(\ggn-\tfrac{1}{4}\cgn\right)\cos \beta   = -\tfrac{1}{2}\cns  \left(\cos^2 \beta-1\right) \label{GLdomB}
\end{align}
when $\theta=\beta+\pi/2$.
Each of the equations \cref{LSdomA,LSdomB,GLdomA,GLdomB} may each be written in terms of just two parameters as follows: \cref{LSdomA,LSdomB} may be written as 
\begin{align}
    \Sn+1-\cos \beta &= -\Dgn \cos \beta \left(\cos^2 \beta-1\right), \label{LSdom} 
\end{align}
while \cref{GLdomA,GLdomB} may be written as
\begin{align}    
    \Sn+1-\cos \beta &= \Dns \left(\cos^2 \beta-1\right), \label{GLdom}
\end{align}
where $\Sn$, $\Dns$ and $\Dgn$ are defined by
\begin{align}
    \Sn &=\dfrac{4\ggs - \left(4\gns-|\cns|\right)}{4\ggn-|\cgn|} - 1, \label{Sn}  \\
    \Dns &=\dfrac{ 2\cns}{4\ggn-|\cgn|}, \label{dells} \\
    \Dgn &=\dfrac{2\cgn}{4\ggn-|\cgn|}, \label{delgl}
\end{align}
respectively.
Note that while the {\it nematic spreading parameter} $\Sn$ is the appropriate generalisation of the isotropic spreading parameter $\Si$ defined in \cref{Si}, the scaled anchoring coefficients $\Dns$ and $\Dgn$ have no isotropic counterparts.
We also note that when $\Dns=\Dgn=0$ (\ie when $\cgn =\cns=0$) then $\Sn=(\ggs-\gns)/\ggn-1=\Si$, and both of the nematic Young equations \cref{LSdom,GLdom} reduce to the classical isotropic Young equation \cref{isoYE}.

Each of the right-hand sides of the nematic Young equations \cref{LSdom,GLdom} involve only one parameter, namely the scaled anchoring coefficients $\Dns$ and $\Dgn$, respectively.
At first sight, it may seem counter-intuitive that $\Dns$ appears in the case of GN-dominant anchoring and $\Dgn$ appears in the case of NS-dominant anchoring. 
However, for GN--dominant anchoring the director is aligned with the preferred director orientation of the gas--nematic interface, and so the corresponding anchoring energy, and therefore the couple on the director, is zero. 
The non-zero contribution to the anchoring energy therefore derives from the breaking of the nematic--substrate interface anchoring. 
The corresponding explanation applies to the NS-dominant case.
The right-hand sides of equations \cref{LSdom,GLdom} may therefore be interpreted  physically as the contribution to the balance of stress at the contact line associated with the breaking of the anchoring on the interface with the weaker anchoring.

\section{The equilibrium states and transitions of a nematic ridge}
\label{sec:nem}

With the director angle determined in regions adjacent to the contact lines, we can now use the nematic Young equations \cref{LSdom,GLdom} to determine the continuous and discontinuous transitions between the $\cw$, $\pw$ and $\cd$ states. 
As \cref{LSdom,GLdom} are cubic and quadratic equations for $\cos\beta$, respectively, they can have up to three real solutions for $\beta$ and up to two real solutions for $\beta$, respectively.
Each of these solutions for $\beta$ corresponds to a different $\pw$ state, and therefore, unlike for the isotropic ridge described in \cref{sec:iso}, a nematic ridge can have multiple $\pw$ states.

Following the same approach as for the isotropic ridge in \cref{sec:iso}, the values of $\Sn$ and $\Dns$ (for GN-dominant anchoring) or $\Sn$ and $\Dgn$ (for NS-dominant anchoring) at which there is a change in the number of possible equilibrium states are again called the transition points.
Specifically, a transition occurs as $\Sn$, $\Dns$ or $\Dgn$ increases or decreases if the previous minimum energy state ceases to exist or a new minimum energy state comes into existence. In an analogous manner to that in the isotropic case, at $\Sn = -2$ and $\Sn = 0$ the number of equilibrium states changes, 
which leads to transitions to a new equilibrium state as $\Sn$ increases or decreases through these values. 
However, unlike in the isotropic case, in which only continuous transitions occur, in the nematic case discontinuous transitions can also now occur, \ie the contact angle can transition discontinuously.

\begin{figure}[tp]
\begin{center}
\begin{tabular}{cc}
 \includegraphics[width=0.47\linewidth]{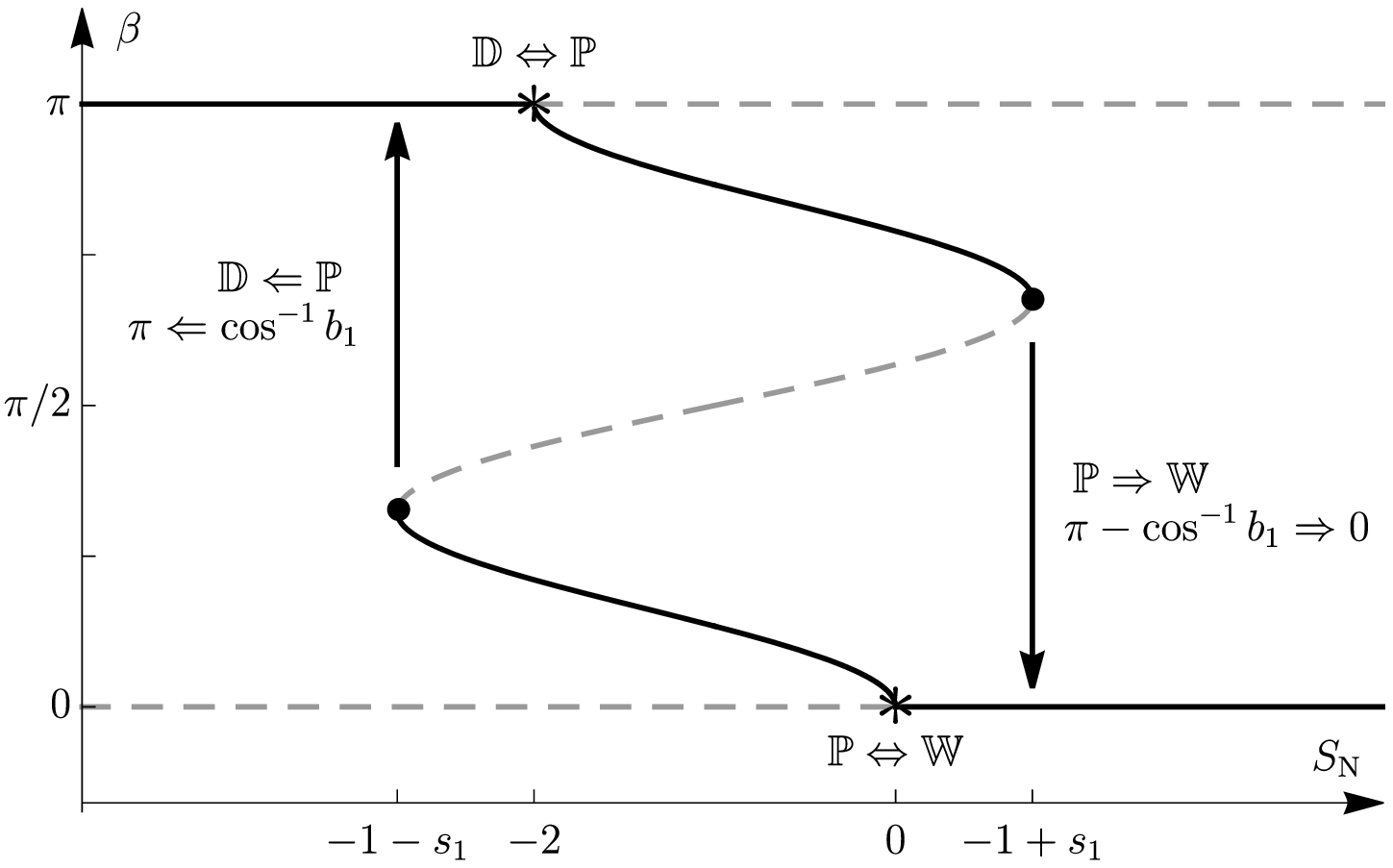} &
 \includegraphics[width=0.47\linewidth]{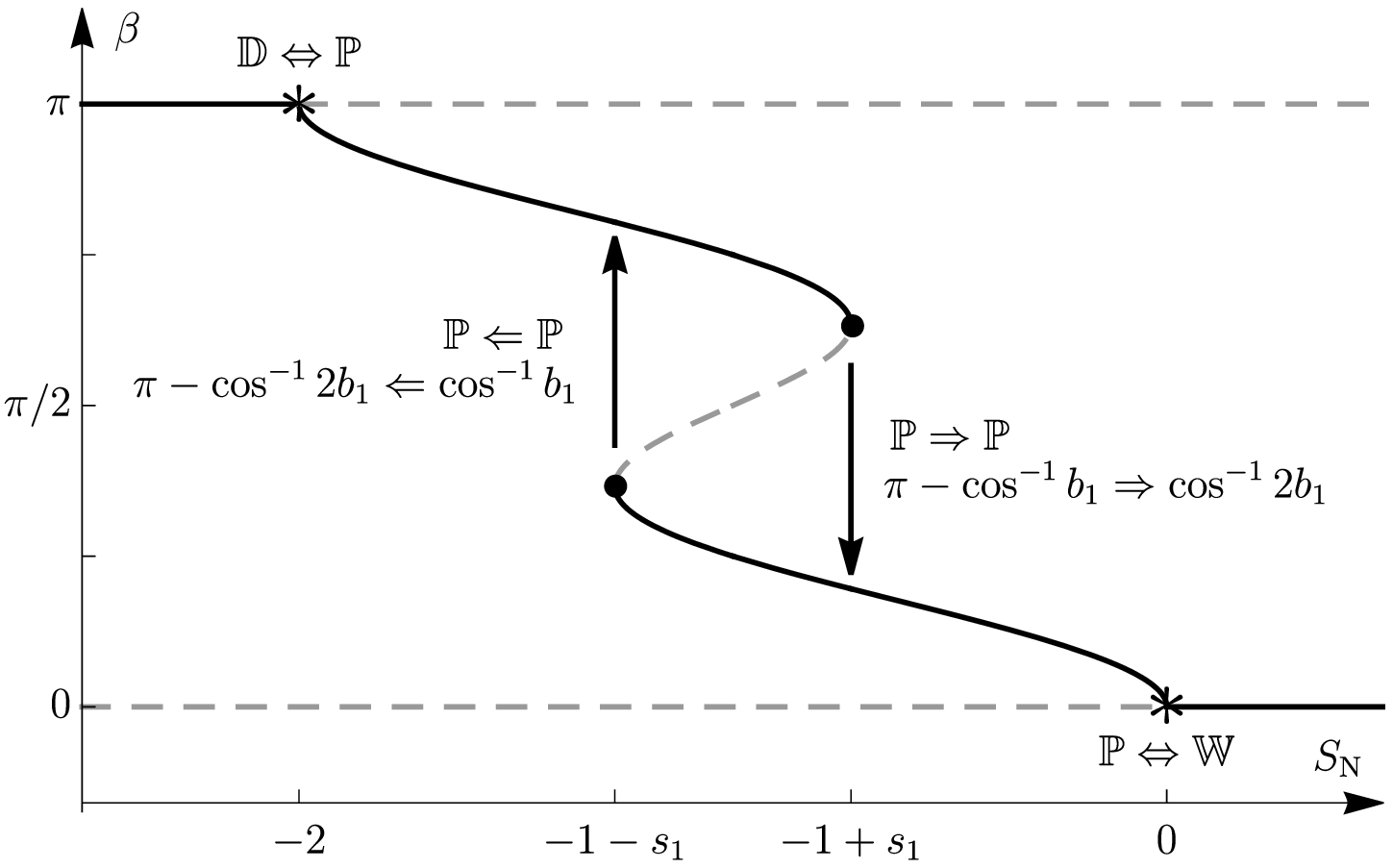} \\
 (a) & (b) \\
 \includegraphics[width=0.47\linewidth]{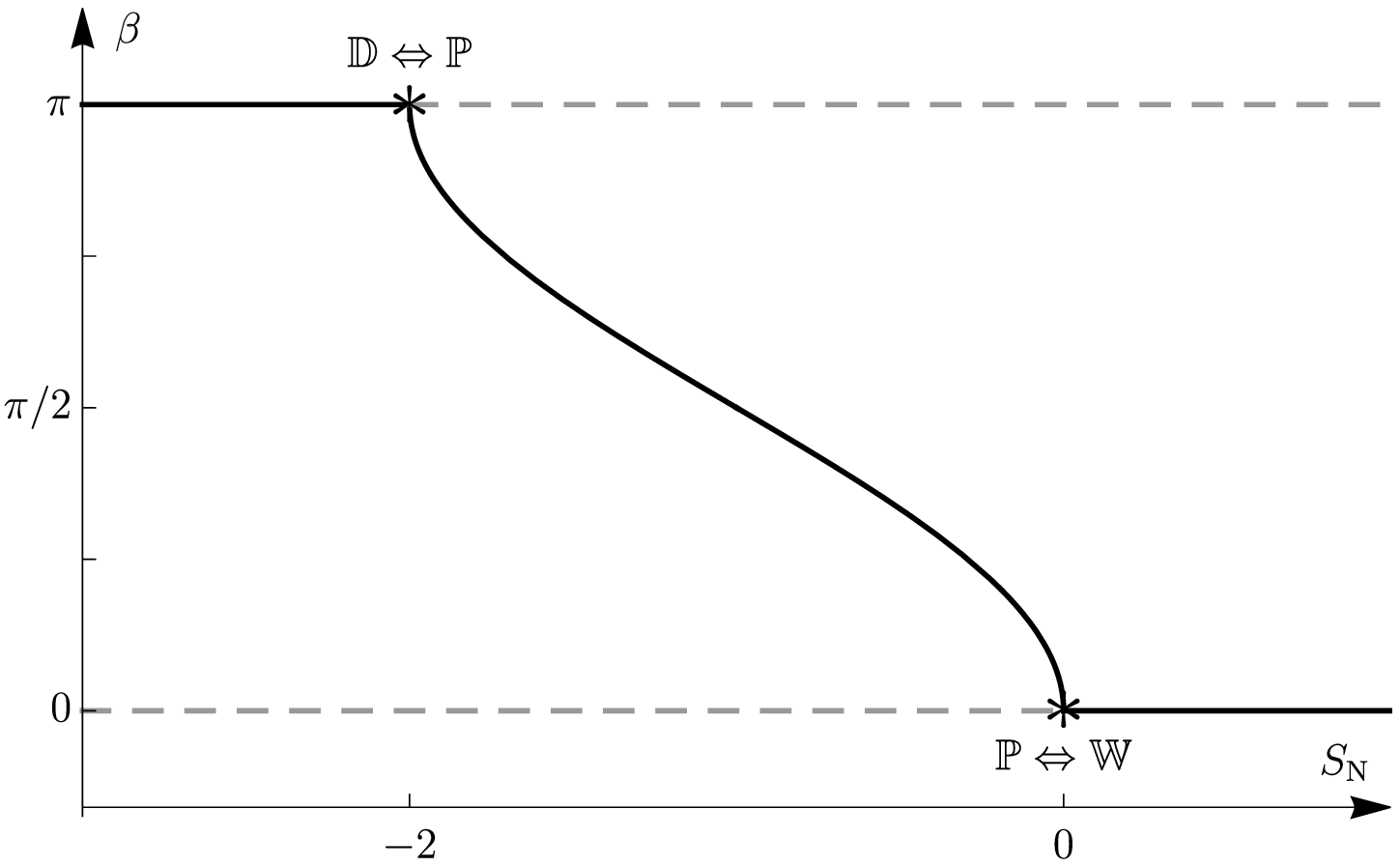} &  
 \includegraphics[width=0.47\linewidth]{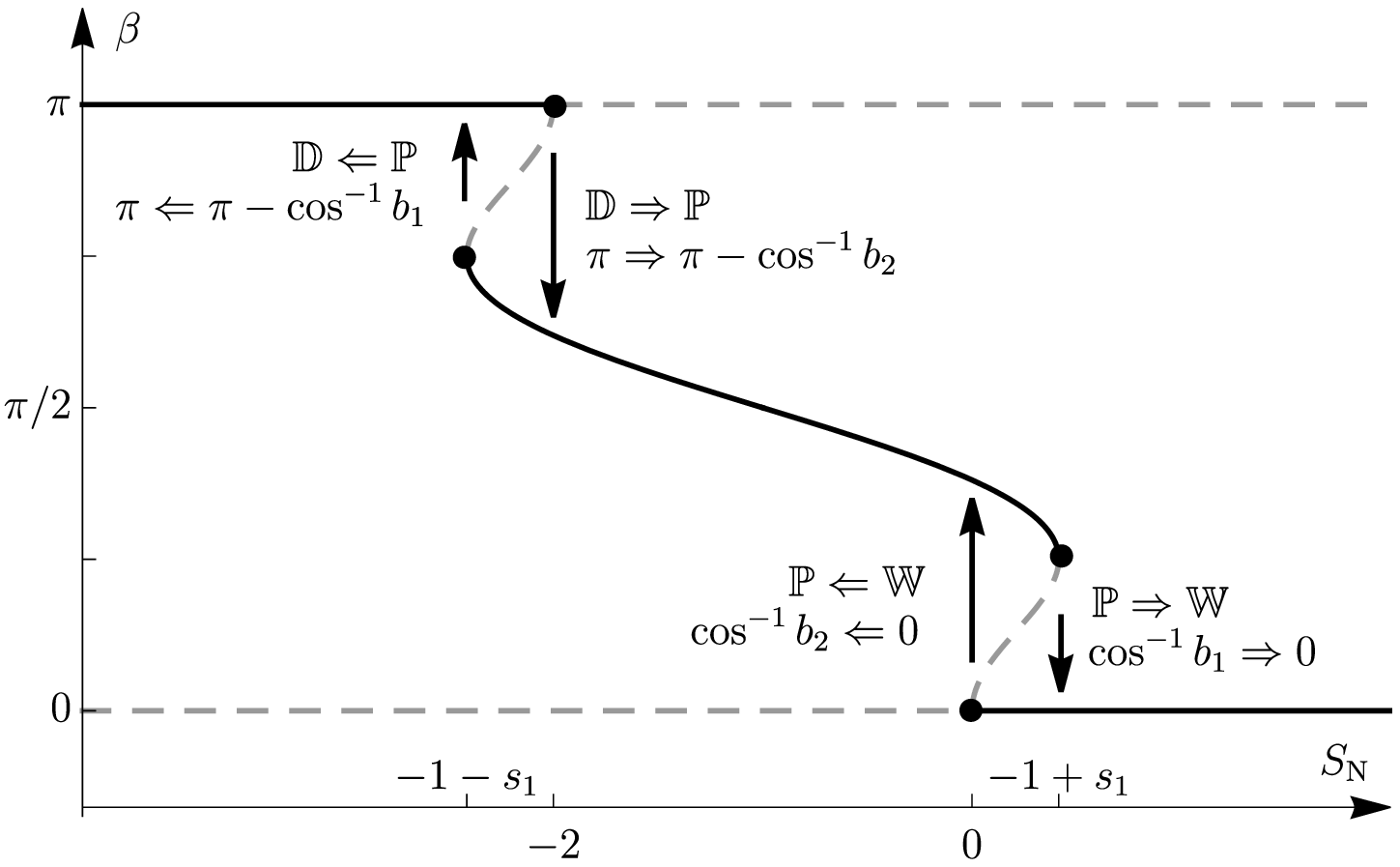} \\
 (c) & (d) \\
\end{tabular}
\end{center}
\caption{
  Summaries of the solutions for the contact angle $\beta$ as a function of the nematic spreading parameter $\Sn$ for NS-dominant anchoring according to \cref{LSdom} for the four ranges of $\Dgn$: (a) $\Dgn<-4$, (b) $-4\le\Dgn\le-1$, (c) $-1\le\Dgn\le1/2$, and (d) $\Dgn>1/2$. 
  The transition points are labelled and shown by stars $(\Ast)$ for a continuous transition and by dots ($\bullet$) for a discontinuous transition, where $s_1=\sqrt{4(1+\Dgn)^3/(27 \Dgn)}$, $b_1 = \sqrt{(1+\Dgn)/(3\Dgn)}$ and $b_2= -1/2+\sqrt{\Dgn(\Dgn+4)}/(2 \Dgn)$.
  The arrows shows the directions of the associated transitions in $\beta$.
  The solid lines denote the hypothesised local minimum energy states and the dashed lines denote the hypothesised local maximum energy states. 
}
\label{bifurcation1} 
\end{figure}

\begin{figure}[tp]
\begin{center}
\begin{tabular}{cc}
 \includegraphics[width=0.475\linewidth]{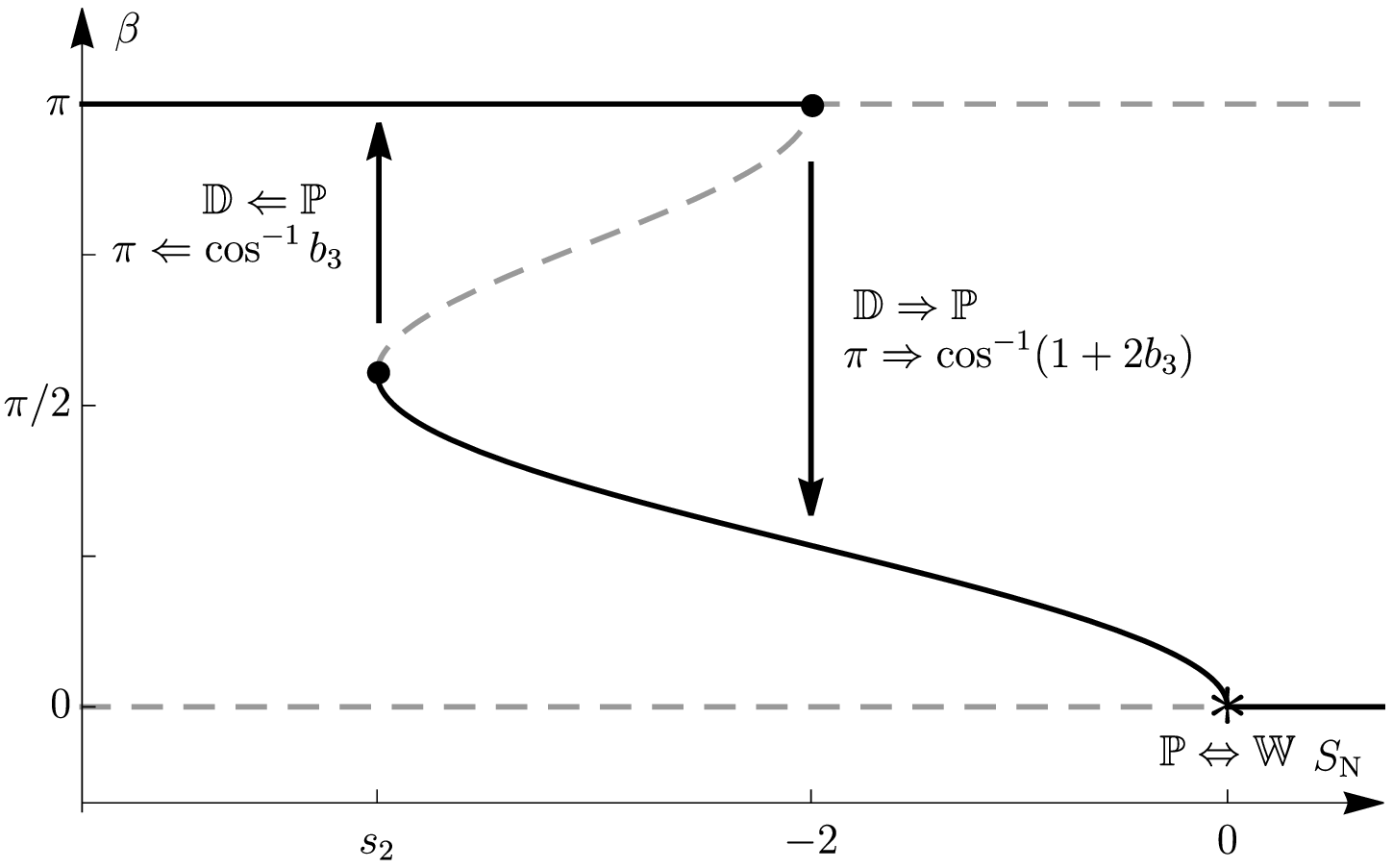} & \includegraphics[width=0.475\linewidth]{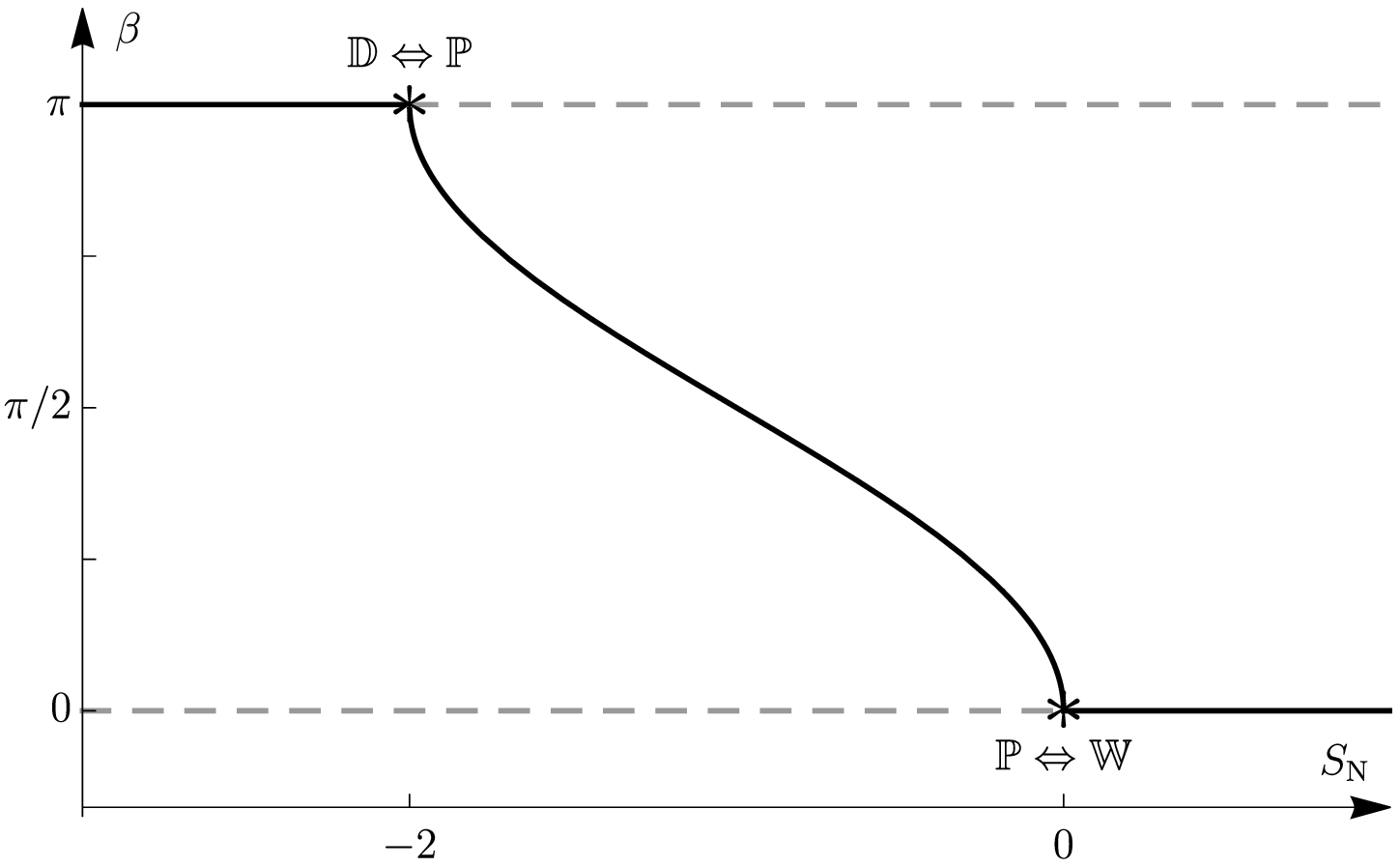} \\
 (a) & (b) \\[0.25cm]
 \multicolumn{2}{c}{\includegraphics[width=0.475\linewidth]{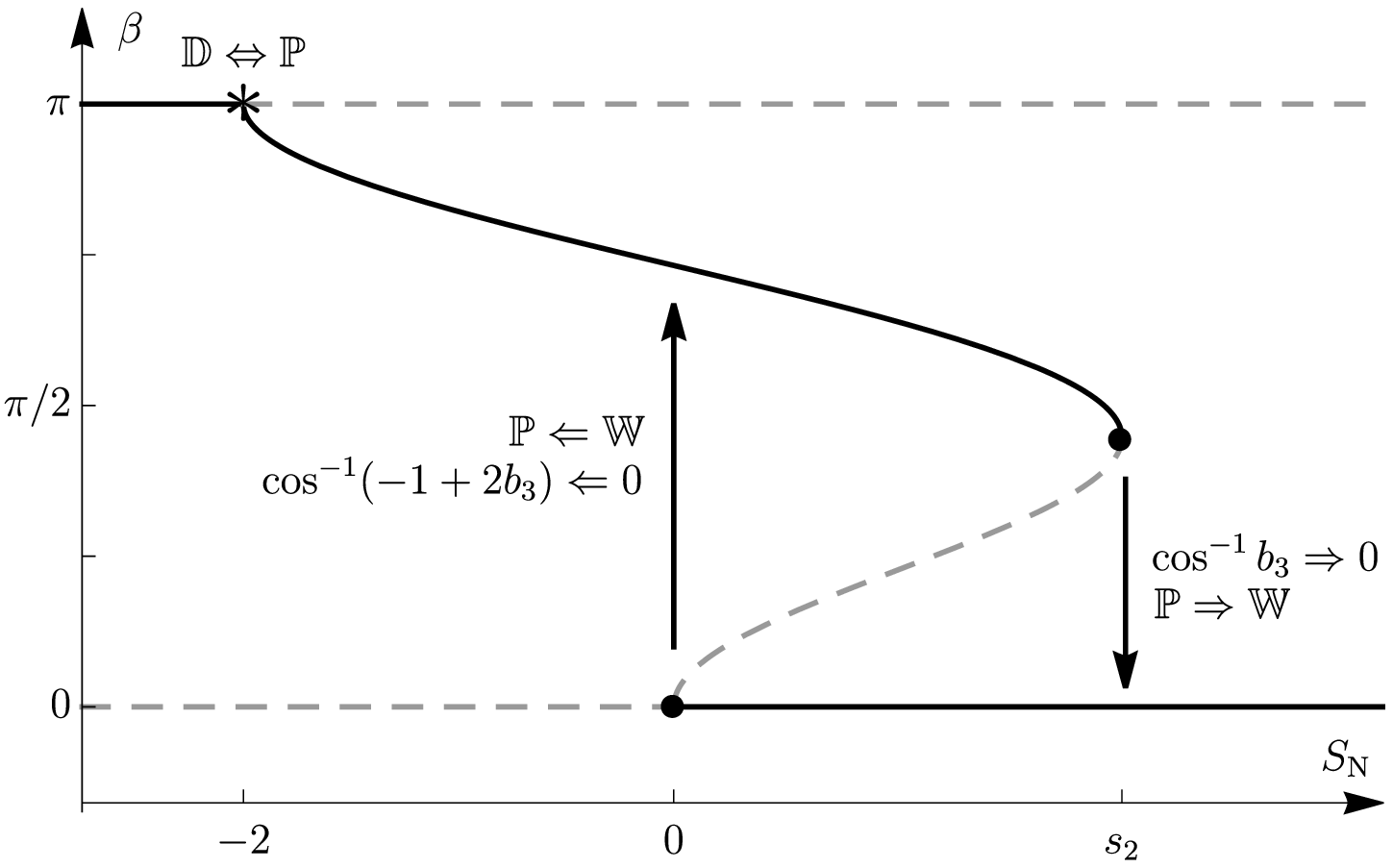}} \\
 \multicolumn{2}{c}{(c)}
\end{tabular}
\end{center}
  \caption{
  Summaries of the solutions for the contact angle $\beta$ as a function of the nematic spreading parameter $\Sn$ for GN-dominant anchoring according to \cref{GLdom} for the three ranges of $\Dns$: (a) $\Dns<-1/2$, (b) $-1/2\le\Dns\le1/2$, and (c) $\Dns>1/2$.
  The transition points are labelled and shown by stars $(\Ast)$ for a continuous transition and by dots ($\bullet$) for a discontinuous transition, where $s_2=-1-\Dns -1/(4 \Dns)$ and $b_3=-1/(2\Dns)$.  
  The arrows shows the directions of the associated transitions in $\beta$.
  The solid lines denote the hypothesised local minimum energy states and the dashed lines denote the hypothesised local maximum energy states. 
  }
  \label{bifurcation2} 
\end{figure}

In both NS-dominant and GN-dominant anchoring, the nature of the different transitions, the contact-angle transitions, and the transition points can be obtained from just the nematic Young equations \cref{LSdom,GLdom}.
In NS-dominant anchoring the transition behaviour depends on whether $\Dgn<-4$, $-4 \le \Dgn <-1$, $-1 \le \Dgn \le 1/2$ or $\Dgn > 1/2$, whereas in GN-dominant anchoring the transition behaviour depends on whether $\Dns<-1/2$, $-1/2 \le \Dns \le 1/2$ or $\Dns > 1/2$.
\cref{bifurcation1,bifurcation2} show summaries of the solutions for the contact angle $\beta$ as a function of the nematic spreading parameter $\Sn$ for these four ranges of $\Dgn$ for NS-dominant anchoring and for these three ranges of $\Dns$ for GN-dominant anchoring, respectively.
%
%
In \cref{bifurcation1,bifurcation2}, and what follows, a rightward arrow ($\Rightarrow$) denotes a discontinuous transition for increasing $\Sn$, and a leftward arrow ($\Leftarrow$) denotes a discontinuous transition for decreasing $\Sn$. 
Thus, for example, a discontinuous transition from complete wetting to partial wetting for increasing $\Sn$ is denoted by $\cw \Rightarrow \pw$, and a discontinuous transition from partial wetting to complete wetting for decreasing $\Sn$ is denoted by $\cw \Leftarrow \pw$.
In addition, we denote a discontinuous transition in the contact angle using the same notation, so that, for example, the contact-angle transition for a $\cw \Rightarrow \pw$ transition, for which the contact angle transitions discontinuously from $\beta=0$ to $\beta=\beta^*$, is denoted by $0\Rightarrow\beta^*$. 
Summaries of all of the possible transitions shown in \cref{bifurcation1,bifurcation2} are given in \cref{tableSls,tableSgl} for NS-dominant and GN-dominant anchoring, respectively.
%
\begin{table}[tp]
\begin{center}
{\renewcommand{\arraystretch}{1.5}
\hspace*{-0.4cm}\begin{tabular}{|c|c|c|c|c|}
\hline
Range of $\Dgn$ & $\Sn$ value at transition &  Nature of the transition & Contact-angle transition \\ \hline
\multirow{4}{*}{$\Dgn < -4$} & $-1 - s_1 \: (<-2)$ & $\cd \Leftarrow \pw$  & $\pi \Leftarrow \cos^{-1} b_1$ \\ \cline{2-4}
                 &  $-2$ & $\cd \Leftrightarrow \pw$ & Continuous with $\beta=\pi$ \\ \cline{2-4}
                 &  $0$ & $\pw \Leftrightarrow \cw$  & Continuous with $\beta=0$ \\ \cline{2-4}
                 & $(0<) \, -1+s_1$ & $\pw \Rightarrow \cw$ & $\pi-\cos^{-1}b_1 \Rightarrow 0$ \\ \hline
\multirow{4}{*}{$-4 \le \Dgn < -1$} & $-2$ &  $\cd \Leftrightarrow \pw$ & Continuous with $\beta=\pi$ \\ \cline{2-4}
                & $(-2\le) \, -1- s_1 \, (<-1)$ & $\pw \Leftarrow \pw$ & $\pi-\cos^{-1}2b_1 \Leftarrow \cos^{-1}b_1$ \\ \cline{2-4}
                 & $(-1<) \, -1 + s_1 \, (\le 0)$ & $\pw \Rightarrow \pw$ & $\pi-\cos^{-1}b_1 \Rightarrow \cos^{-1}2b_1$ \\ \cline{2-4} 
                & $0$ & $\pw \Leftrightarrow \cw$ & Continuous with $\beta=0$ \\ \hline
\multirow{2}{*}{$-1 \le \Dgn \le 1/2$} & $-2$ & $\cd \Leftrightarrow \pw$ & Continuous with $\beta=\pi$ \\ \cline{2-4}
               & $0$ & $\pw \Leftrightarrow \cw$  & Continuous with $\beta=0$ \\ \hline
\multirow{4}{*}{$\Dgn>1/2$} & $-1- s_1 \, (<-2)$ & $\cd \Leftarrow \pw$ & $\pi \Leftarrow \pi-\cos^{-1} b_1$ \\ \cline{2-4}
                  & $-2$ & $\cd \Rightarrow \pw$  & $\pi \Rightarrow \pi-\cos^{-1} b_2$  \\ \cline{2-4} 
                  & $0$ &  $\pw \Leftarrow \cw$  & $\cos^{-1} b_2 \Leftarrow 0$\\ \cline{2-4} 
                  & $(0<) \, -1+s_1$ & $\pw \Rightarrow \cw$  &$\cos^{-1} b_1 \Rightarrow 0$ \\ \hline
\end{tabular}}
\end{center}
\caption{
Summary of all of the possible transitions for NS-dominant anchoring obtained from \cref{LSdom}.
The four ranges of values of $\Dgn$, the $\Sn$ value at which transitions occur, where $s_1=\sqrt{4(1+\Dgn)^3/(27 \Dgn)}$, and the nature of the different transitions and the contact-angle transitions, where $b_1 = \sqrt{(1+\Dgn)/(3\Dgn)}$ and $b_2= -1/2+\sqrt{\Dgn(\Dgn+4)}/(2 \Dgn)$, are shown.
}
\label{tableSls}
\end{table}
%
\begin{table}[tp]
\begin{center}
{\renewcommand{\arraystretch}{1.5}
\hspace*{-0.4cm}\begin{tabular}{|c|c|c|c|}
        \hline
        Range of $\Dns$ & $\Sn$ value at transition & Nature of the transition & Contact-angle transition \\ \hline
        \multirow{3}{*}{$\Dns < -1/2$} 
        & $-2$ & $\cd \Leftrightarrow \pw$  & Continuous with $\beta=\pi$ \\ \cline{2-4} 
        & $0$ & $\pw \Leftarrow \cw$  & $\cos^{-1} \left(-1 +2b_3 \right) \Leftarrow 0$ \\ \cline{2-4} 
        &  $(0<) \, s_2$ & $\pw \Rightarrow \cw$  & $\cos^{-1} b_3 \Rightarrow 0$ \\ \hline
        \multirow{2}{*}{$-1/2 \le \Dns \le 1/2$} 
        & $-2$ & $\cd \Leftrightarrow \pw$  & Continuous with $\beta=\pi$ \\ \cline{2-4}
        & $0$ & $\pw \Leftrightarrow \cw$ & Continuous with $\beta=0$ \\ \hline
        \multirow{3}{*}{$\Dns> 1/2$} 
        & $s_2\, (<-2)$ & $\cd \Leftarrow \pw$  & $\pi \Leftarrow \cos^{-1} b_3$ \\ \cline{2-4}
        & $-2$ & $\cd \Rightarrow \pw$  & $\pi \Rightarrow \cos^{-1} \left( 1 +2b_3\right)$ \\ \cline{2-4}
        &  $0$ & $\pw \Leftrightarrow \cw$  & Continuous with $\beta=0$ \\ \hline
\end{tabular}}
\end{center}
\caption{
Summary of all of the possible  transitions for GN-dominant anchoring obtained from \cref{GLdom}.
The three ranges of values of $\Dgn$, the $\Sn$ value at which transitions occur, where $s_2=-1-\Dns -1/(4 \Dns)$, and the nature of the different transitions and the contact-angle transitions, where $b_3=-1/(2\Dns)$, are shown.
}
\label{tableSgl}
\end{table}

Although for a nematic ridge, unlike for an isotropic ridge, the free energy of each equilibrium state cannot be determined analytically, we can hypothesise the local minimum energy states for the nematic ridge by comparison with those for the isotropic ridge described in \cref{sec:iso}.
Hence we hypothesise that the $\cd$ state is a local minimum energy state for $\Sn<-2$ and a local maximum energy state for $\Sn\ge-2$.
Similarly, we hypothesise that the $\cw$ state is a local minimum energy state for $\Sn > 0$ and a local maximum energy state for $\Sn\le0$.
Assuming that there will always be at least one local minimum energy state, within the range $-2 \le \Sn \le 0$, where the $\cw$ and the $\cd$ states are local maximum energy states, the local minimum energy state must be a $\pw$ state.
The local minimum and maximum energy states are shown in \cref{bifurcation1,bifurcation2} by solid lines and dashed lines, respectively.
In the absence of a full dynamical theory, we also hypothesise that the continuous and discontinuous transitions shown in \cref{bifurcation1,bifurcation2} each correspond to classical pitchfork or fold bifurcations \cite{GuckenheimerHolmes}.
In particular, the transitions at $\Sn=-2$ and $\Sn=0$ are pitchfork bifurcations, where a change in $\Sn$, $\Dgn$ or $\Dns$ leads to a local minimum energy state becoming a local maximum energy state, forcing the system to transition continuously (through a super-critical pitchfork bifurcation) or discontinuously (through a sub-critical pitchfork bifurcation) to a new local minimum energy state.
Furthermore, the discontinuous transitions at $\Sn=-1\pm s_1$ and $\Sn=s_2$, where 
\begin{align}
    s_1&=\sqrt{4(1+\Dgn)^3/(27 \Dgn)},\\
    s_2&=-1-\Dns-1/(2\Dns),
\end{align}
are associated with fold bifurcations, where a change in $\Sn$, $\Dgn$ or $\Dns$ leads to a local minimum energy state combining with a local maximum energy state, forcing the system to transition discontinuously to a different local minimum energy state.

Figures \ref{bifurcation1}(a) and (b) and Figures \ref{bifurcation2}(a) and (c) also show that there are ranges of $\Sn$ values for which there are two local minimum energy states (shown by solid lines).
Perhaps most interestingly, we see from \cref{bifurcation1}(a) that when $-2 \le \Sn \le 0$ and from \cref{bifurcation1}(b) that when \mbox{$-1-s_1 \le \Sn \le -2$} there are two local minimum energy $\pw$ states.
This implies that the effects of anchoring breaking can give rise to two local minimum energy $\pw$ states, a situation that does not occur in the isotropic case.

From the results summarised in \cref{tableSgl,tableSls} the asymptotic behaviour of the contact-angle transitions in the limits of large  anchoring coefficients relative to the isotropic interfacial tension, namely the limits  $\Dgn \to \pm \infty$ and $\Dns \to \pm \infty$, may be determined.
For example, for NS-dominant anchoring, as $\Dgn\to\infty$ the contact-angle transition for the $\pw \Leftarrow \cw$ transition approaches a discontinuous transition in contact angle from $\beta=0$ to $\beta=\pi/2$, and the contact-angle transition for the $\cd \Rightarrow \pw$ transition approaches a discontinuous transition in the contact angle from $\beta=\pi$ to $\beta=\pi/2$.
This limiting behaviour shows that for GN-dominant anchoring in the limit $\Dns\to \infty$ the contact-angle transition for the $\pw \Leftarrow \cw$ transition approaches a discontinuous transition in the contact angle from $\beta=0$ to $\beta=\pi$, \ie it approaches a discontinuous transition from the $\cw$ state directly to the $\cd$ state, which bypasses the $\pw$ state. Similarly, in the limit $\Dns\to -\infty$ the contact-angle transition for the $\cd \Rightarrow \pw$ transition approaches a discontinuous transition in the contact angle from $\beta=\pi$ to $\beta=0$, \ie it approaches a discontinuous transition from the $\cd$ state directly to the $\cw$ state.

The discontinuous transitions shown in Figures \ref{bifurcation1}(a), (b), and (d) and Figures \ref{bifurcation2}(a) and (c) show that the behaviour of the contact angle is hysteretic. This nematic contact-angle hysteresis, which occurs for an ideal substrate, is fundamentally different from the well-known phenomenon of isotropic contact-angle hysteresis which, as we have previously mentioned, occurs only for a non-ideal substrate.
However, we note that when $-1\le\Dgn\le1/2$ for NS-dominant anchoring, as shown in \cref{bifurcation1}(c), and when $-1/2 \le \Dns \le 1/2$ for GN-dominant anchoring, as shown in \cref{bifurcation2}(b), the behaviour is similar to the isotropic case and no contact-angle hysteresis occurs.

\section{Conclusions}

In the present work we analysed a two-dimensional static ridge of nematic resting on an ideal solid substrate in an atmosphere of passive gas.
In \cref{sec:model,sec:CoV}, we obtained the first complete theoretical description for this system by minimising the free energy, which is given by the sum of the bulk elastic energy, gravitational potential energy and the interface energies, subject to a prescribed constant cross-sectional area. 
We then, in \cref{sec:GE}, chose explicit forms of the bulk elastic energy density and the interface energy densities, namely the standard Oseen--Frank bulk elastic energy density and the standard Rapini--Papoular interface energy densities, and obtained the governing equations \cref{EqBulkF,EqGLF,EqLSF,EqYLF,EqYEF1,EqYEF2}.
Specifically, these equations determine the director angle $\theta(x,z)$, the ridge height $h(x)$, the contact line positions $x=\dpn$, and the Lagrange multiplier $p_0$, in terms of the physical parameters, namely the splay and bend elastic constants $K_1$ and $K_3$, the corresponding isotropic interfacial tensions $\ggn$, $\gns$ and $\ggs$, and the anchoring strengths $\cgn$ and $\cns$.
These governing equations can, in principle, be generalised to include electromagnetic forces, additional contact line effects, non-ideal substrates, or more detailed models for the nematic molecular order, such as Q-tensor theory \cite{Qtensor}.

After briefly reviewing the behaviour of an isotropic ridge in \cref{sec:iso} and discussing the nematic Young equations \cref{EqYEF1,EqYEF2} in \cref{sec:nemYE}, in \cref{sec:nem}, under the assumption that anchoring breaking occurs in regions adjacent to the contact lines, we used the nematic Young equations \cref{EqYEF1,EqYEF2} to determine the continuous and discontinuous transitions that occur between the $\cw$, $\pw$ and $\cd$ states.
In particular, it was shown that the nematic Young equations in the cases of NS-dominant and GN-dominant anchoring, which are given by \cref{LSdom,GLdom}, respectively, can each be written in terms of two parameters, namely the nematic spreading parameter $\Sn$ and one of the scaled anchoring coefficients $\Dgn$ and $\Dns$.
In both situations, we found continuous transitions analogous to those that occur in the classical case of an isotropic liquid, but also a variety of discontinuous transitions, as well as contact-angle hysteresis, and regions of parameter space in which there exist multiple partial wetting states that do not occur in the classical case of an isotropic liquid.
Summaries of all the transitions for NS-dominant and GN-dominant anchoring are given in \cref{bifurcation1,bifurcation2}, respectively, and in \cref{tableSls,tableSgl}, respectively.

For simplicity, in \cref{sec:nem} we considered only the left-hand contact line, which is described by the nematic Young equation \cref{EqYEF1}. Corresponding results can be obtained for the right-hand contact line, and, since we have shown that  there is more than one possible contact angle value for the same parameter values, $\betaLR$ do not, in general, have to be the same and so the ridge can be asymmetric.
This is in agreement with observations by Vanzo \etal \cite{Vanzo2016}, who found that anisotropic effects can lead to multiple contact-angle values and asymmetry of  elongated sessile nematic droplets.

Concerning potential future comparisons with the results of physical experiments of the situation modelled in the present work, we have shown that discontinuous transitions and contact-angle hysteresis will occur if the parameters are such that $\Dgn>1/2$ or $\Dgn<-1$, or $|\Dns|>1/2$.
Inspection of \cref{delgl,dells} shows that one of these inequalities may be satisfied  when the isotropic interfacial tension and the anchoring strength are of similar magnitude at one of the interfaces, \ie $\ggn \simeq |\cgn|$ or $\gns \simeq |\cns|$. 
For standard low-molecular-mass nematics, for which the isotropic interfacial tension is typically larger than the anchoring strength \cite{SONINBOOK}, this may be difficult to achieve. 
For example, for the typical parameter values given in \cref{sec:GE}, 
$|\Dgn| \ll 1$ and $|\Dns| \ll 1$. Therefore,  the present analysis indicates that, as many previous authors have implicitly or explicitly assumed, for standard low-molecular-mass nematics the classical isotropic Young equations \cref{isoYE1} are a good approximation for the nematic Young equations \cref{EqYEF1,EqYEF2} and discontinuous  transitions and contact-angle hysteresis will not be observed. However, for high-molecular-mass nematics, \eg nematic polymers, or systems with particularly strong anchoring, the anchoring strengths would be considerably higher, and the discontinuous transitions could potentially be observed experimentally. 
For example, the use of polymeric compounds to produce tailored anchoring \cite{vanderWielen2000} leads to a strong preference for polymers to align at interfaces \cite{Zhouetal2005,Lietal2018} and may result in large anchoring strengths, which could lead to $|\Dgn| =O(1)$ and $|\Dns| = O(1)$ and hence the transitions predicted in the present work could potentially be observed.
Alternatively, the situation in which the surrounding fluid is the isotropic melt of the nematic could lower the isotropic interfacial tension $\ggn$. 
In this situation, the isotropic interfacial tension for the isotropic--nematic interface $\gin$ would be much smaller than the gas--nematic interfacial tension $\ggn$ and may become comparable with the anchoring strength $\cgn$. 
For instance, $\gin$ was measured for the nematic MBBA as $\gin=10^{-5}\,{\rm N m}^{-1}$ \cite{LangevinBouchiat1973}, which is three orders of magnitude smaller than typical isotropic interfacial tension for a gas--nematic interface $\ggn$ \cite{Dhara2020}.
Such a situation could be realised experimentally by using controlled heating and cooling of regions of a substrate coated in a nematic film \cite{vanderWielen2000,DharaMukherjee2019}. 

The range of anisotropic wetting and dewetting phenomena occurring in this nematic system may also be useful from a technological perspective, for instance, for tailored dewetting of liquid films, as discussed in \cref{sec:Intro}  \cite{Gentili2012,Bramble2010,Zou2018,Blow2013,Brown2009}. The variety of possible transitions between two-dimensional equilibrium states will have similar forms in three dimensions, which may be relevant to applications such as the One Drop Filling method of LCD manufacturing \cite{LowErCousins,Cousins2020,OneDropFill1} and adaptive-lens technologies \cite{Algorri2019,Kim2017}. In order to explore such applications, further theoretical investigations, particularly into the dynamics of transitions, and experimental investigations would be needed.
\bibliography{main}
\end{document}